\documentclass[11pt,a4paper]{article}
\usepackage[utf8]{inputenc}
\usepackage[T1]{fontenc} 
\usepackage[english]{babel}
\usepackage[top=2.5cm, bottom=3.0cm, left=2.5cm, right=2.5cm]{geometry}
\usepackage{setspace}
\usepackage{dutchcal}
\makeatletter
\g@addto@macro\bfseries{\boldmath}
\makeatother
\usepackage{amssymb}
\usepackage{amsmath}
\usepackage{amsfonts}
\usepackage{mathrsfs}
\usepackage{bm}
\usepackage{slashed}
\numberwithin{equation}{section}
\allowdisplaybreaks
\usepackage{enumitem}
\usepackage{array}

\usepackage[]{caption} 
\usepackage{hyperref}
\hypersetup{
    colorlinks,%
    citecolor=blue,%
    filecolor=blue,%
    linkcolor=blue,%
   	urlcolor=blue,
   	linktoc=page
}
\usepackage{cite}
\usepackage[toc,page]{appendix}
\usepackage{graphicx}
\usepackage{float}
\usepackage{subfig}
\usepackage{tikz}
\usetikzlibrary{calc}
\usepackage{xcolor}
\definecolor{DarkGreen}{RGB}{0,128,0}
\usepackage{mdframed}
\newcommand{\p}{\partial}

\newcommand{\ndelta}{\delta\hspace{-0.50em}\slash\hspace{-0.05em} }
\newcommand{\poubelle}[1]{}

\renewcommand{\textbf}[1]{\begingroup\bfseries\mathversion{bold}#1\endgroup}
\newcommand{\D}{\text d}

\usepackage{pict2e}
\makeatletter
\newcommand{\loplus}{\mathbin{\mathpalette\dog@lsemi{+}}}
\newcommand{\dog@lsemi}[2]{\dog@semi{#1}{#2}{270,90}}
\newcommand{\dog@semi}[3]{%
  \begingroup
  \sbox\z@{$\m@th#1#2$}%
  \setlength{\unitlength}{\dimexpr\ht\z@+\dp\z@\relax}%
  \makebox[\wd\z@]{\raisebox{-\dp\z@}{%
    \begin{picture}(1,1)
    \linethickness{\variable@rule{#1}}
    \roundcap
    \put(0.5,0.5){\makebox(0,0){\raisebox{\dp\z@}{$\m@th#1#2$}}}
    \put(0.5,0.5){\arc[#3]{0.5}}
    \end{picture}%
  }}%
  \endgroup
}
\newcommand{\variable@rule}[1]{%
  \fontdimen8  
  \ifx#1\displaystyle\textfont3\else
    \ifx#1\textstyle\textfont3\else
      \ifx#1\scriptstyle\scriptfont3\else
        \scriptscriptfont3\relax
  \fi\fi\fi
}
\makeatother

\begin{document}

\setstretch{1.1}

\setcounter{tocdepth}{2}

\begin{titlepage}

\begin{flushright}\vspace{-3cm}
{\small
\phantom{\today} }\end{flushright}

\begin{center}
\setstretch{1.75}
{\LARGE{\textbf{ 
The $\Lambda$-BMS$_4$ Charge Algebra
}}}
\end{center}
 \vspace{7mm}
\begin{center} 
\centerline{\large{\bf{Geoffrey Comp\`{e}re\footnote{e-mail: gcompere@ulb.ac.be}, Adrien Fiorucci\footnote{e-mail: afiorucc@ulb.ac.be}, Romain Ruzziconi\footnote{e-mail: rruzzico@ulb.ac.be}}}}

\vspace{2mm}
\normalsize
\bigskip\medskip
\textit{Universit\'{e} Libre de Bruxelles and International Solvay Institutes\\
CP 231, B-1050 Brussels, Belgium\\
\vspace{2mm}
}
\vspace{10mm}

\begin{abstract}
The surface charge algebra of generic asymptotically locally (A)dS$_4$ spacetimes without matter is derived without assuming any boundary conditions. Surface charges associated with Weyl rescalings are vanishing while the boundary diffeomorphism charge algebra is non-trivially represented without central extension. The $\Lambda$-BMS$_4$ charge algebra is obtained after specifying a boundary foliation and a boundary measure. The existence of the flat limit requires the addition of corner terms in the action and symplectic structure that are defined from the boundary foliation and measure. The flat limit then reproduces the BMS$_4$ charge algebra of supertranslations and super-Lorentz transformations acting on asymptotically locally flat spacetimes. The BMS$_4$ surface charges represent the BMS$_4$ algebra without central extension at the corners of null infinity under the standard Dirac bracket, which implies that the BMS$_4$ flux algebra admits no non-trivial central extension. \\[1.5cm]


\end{abstract}


\end{center}

\end{titlepage}

\newpage

\begin{spacing}{0.8}

\tableofcontents

\end{spacing}

\newpage

\section{Introduction}
\label{sec:intro}

In asymptotically flat Einstein gravity, leading \cite{PhysRevD.28.1894,He:2014laa,Strominger:2014pwa}, subleading/overleading \cite{Kapec:2014opa,Campiglia:2014yka,Campiglia:2015yka,Pasterski:2015tva,Conde:2016rom,Compere:2018ylh,Distler:2018rwu} or higher subleading/overleading \cite{Hamada:2018vrw,Compere:2019odm} relationships have been found in the infrared sector, which relate symmetries \cite{Bondi:1962px,Sachs:1962wk,Tamburino:1966zz,deBoer:2003vf,Barnich:2009se,Barnich:2010eb,Barnich:2011mi,Campiglia:2014yka,Campiglia:2015yka,Flanagan:2015pxa,Compere:2017wrj,Compere:2018ylh}, memory effects \cite{Zeldovich:1974gvh,1977ApJ...216..610T,1978ApJ...220.1107T,Turner:1978aa,PhysRevD.44.R2945,Blanchet:1987wq,Blanchet:1992br,Christodoulou:1991cr,Thorne:1992sdb,Bieri:2010tq,Pasterski:2015tva,Lasky:2016knh,Nichols:2017rqr,Nichols:2018qac,Flanagan:2018yzh,Compere:2019odm} and scattering identities involving soft modes \cite{Weinberg:1965nx,Cachazo:2014fwa}. Partial reviews include \cite{Ashtekar:2014zsa,Strominger:2017zoo,Compere:2018aar,Ruzziconi:2019pzd}. In cosmology, such relationships have also been described \cite{Hinterbichler:2013dpa,Berezhiani:2013ewa,Horn:2014rta,Mirbabayi:2016xvc,Kehagias:2016zry , Hamada:2017gdg} which relate symmetries \cite{Anninos:2010zf,Hinterbichler:2013dpa,Horn:2014rta,Compere:2019bua,Ashtekar:2014zfa}, memory effects \cite{Bieri:2015jwa , Chu:2015yua, Tolish:2016ggo , Chu:2016qxp , Chu:2016ngc , Hamada:2017gdg,Bieri:2017vni, Chu:2019ssw} and consistency conditions for correlation functions \cite{Maldacena:2002vr,Creminelli:2004yq,Cheung:2007sv,Hinterbichler:2013dpa,Horn:2014rta,Hui:2018cag}. In this context, the main motivation of this work is to increase the understanding of symmetries in cosmological spacetimes with the aim to better understand their infrared structure. In addition, the flat limit of cosmological spacetimes allows for an improved understanding of the infrared structure of asymptotically flat spacetimes. 

In our previous work \cite{Compere:2019bua}, we found how to embed the (often called generalized) BMS$_4$ symmetry algebra \cite{Barnich:2010eb,Barnich:2011ct,Barnich:2011mi,Campiglia:2014yka,Campiglia:2015yka} within asymptotically locally anti-de Sitter or de Sitter spacetimes in four-dimensional Einstein gravity. Two ingredients are necessary: a \emph{foliation} of the codimension one asymptotic boundary manifold, and a \emph{measure} on the codimension two manifolds induced by the foliation. This background structure suffices to uniquely identify the $\Lambda$-BMS$_4$ algebra\footnote{Technically, it is an algebroid \cite{2000math.....12106F,Lyakhovich:2004kr,Barnich:2010eb,Barnich:2010xq,Barnich:2017ubf} (also called soft gauge algebra \cite{Sohnius:1982rs}) instead of a Lie algebra because its structure constants depend upon the boundary metric. We will still use the terminology ``algebra'' for simplicity.}. In this paper, we will further derive the surface charge algebra associated with the $\Lambda$-BMS$_4$ algebra and match the BMS$_4$ charge algebra \cite{Barnich:2011mi,Compere:2018ylh} in its flat limit.

The $\Lambda$-BMS$_4$ algebra arises in asymptotically locally de Sitter spacetimes by fixing a boundary gauge without constraining the Cauchy problem. While the entire boundary diffeomorphism algebra acts on the boundary metric \cite{Anninos:2010zf}, its $\Lambda$-BMS$_4$ sub-algebra captures the entire set of charges that are constrained by Einstein's equations, namely the Bondi mass and Bondi angular momentum, which can be identified as the components of the holographic stress-tensor along the foliation. In locally anti-de Sitter spacetimes, the $\Lambda$-BMS$_4$ algebra also arises in the nAdS$_4$/nCFT$_3$ correspondence (the generalization to four dimensions of the nAdS$_2$/nCFT$_1$ correspondence \cite{Almheiri:2014cka,Maldacena:2016upp}) where AdS$_4$ is coupled to an external system with which it can exchange energy. Instead, in the AdS/CFT correspondence, Dirichlet boundary conditions are imposed and break the $\Lambda$-BMS$_4$ algebra to the standard $SO(3,2)$ algebra \cite{Henneaux:1985tv}. 

In order to  achieve a flat limit of the $\Lambda$-BMS$_4$ surface charge algebra, one new ingredient will be required. We will need to sophisticate the technical definitions of the variational principle and the symplectic structure in the presence of asymptotic boundaries with fluxes, in the tradition of covariant phase space methods \cite{Regge:1974zd,Lee:1990nz, Wald:1993nt, Iyer:1994ys, Wald:1999wa,Barnich:2001jy,Barnich:2007bf,Compere:2018aar,Compere:2008us}. More precisely, we will need to complete the prescription given in \cite{Compere:2008us} to take into account corner terms in the variational principle in addition to the holographic counterterms \cite{Balasubramanian:1999re,deHaro:2000vlm,Skenderis:2002wp}. As a result, we will find the explicit realization of the full BMS$_4$ asymptotic symmetry group under the standard Dirac bracket and prove that it does not admit central extensions at the corners of null infinity.

The rest of the paper is organized as follows. In Section \ref{Charges of asymptotically locally}, we derive the renormalized symplectic structure of generic asymptotically locally (A)dS$_4$ spacetimes using the holographic renormalization procedure of the action \cite{Balasubramanian:1999re,deHaro:2000vlm,Skenderis:2002wp} and of the symplectic structure \cite{Compere:2008us}. We derive the surface charges associated to residual gauge transformations in Starobinsky/Fefferman-Graham gauge \cite{Starobinsky:1982mr,Fefferman:1985aa} and show that their algebra closes without central extension under the adjusted bracket prescribed in \cite{Barnich:2011mi}. In Section \ref{sec:Lambda-BMS charge algebra}, we particularize our general results to obtain the $\Lambda$-BMS$_4$ charge algebra after imposing boundary Dirichlet gauge defined from a boundary foliation and measure. In Section \ref{Flat limit of}, we show that the flat limit of the phase space associated with $\Lambda$-BMS$_4$ symmetry leads to the (generalized) BMS$_4$ phase space derived in \cite{Compere:2018ylh}. In order for this limit to be smooth, we introduce corner terms in the variational principle and we prescribe corresponding boundary terms in the symplectic structure. We finally describe the BMS$_4$ surface charge algebra and its corresponding flux algebra under the adjusted Lie bracket \cite{Barnich:2011mi} and under the standard Dirac bracket. We end with a discussion in Section \ref{sec:Discussion}. 

The main text is supplemented with several appendices in order to make the paper self-contained. In Appendix \ref{app:Lift}, we set our notations and conventions. In Appendix \ref{Details of Al(A)dS gravity}, we discuss Starobinsky/Fefferman-Graham and Bondi gauges in asymptotically locally (A)dS$_4$ spacetimes. Furthermore, we provide some computational details on the holographic renormalization procedure, the surface charges, the adjusted Lie bracket and the adjusted Dirac bracket. In Appendix \ref{app:LBMS4gen}, we provide a derivation of the explicit $\Lambda$-BMS$_4$ generators. Finally, in Appendices \ref{Useful relations} and \ref{Flat limit of solution space and symmetries}, we summarize useful results obtained in \cite{Compere:2019bua} concerning the flat limit.

\section{Charge algebra of asymptotically locally (A)dS$_{\mathbf 4}$ spacetimes}
\label{Charges of asymptotically locally}

In this section, we first review the well-known holographically renormalized action for asymptotically locally (A)dS$_4$  spacetimes (Al(A)dS$_4$ spacetimes in short) in Einstein gravity without matter. We then review how the action leads to a renormalized symplectic structure and we derive the covariant Hamiltonian associated to all residual diffeomorphisms of Starobinsky/ Fefferman-Graham gauge, namely the Weyl rescalings and boundary diffeomorphisms. We will prove from first principles that the surface charges associated with Weyl rescalings are vanishing and therefore do not belong to the asymptotic symmetry group. Instead, the boundary diffeomorphisms admit non-vanishing Hamiltonian surface charges and do belong to the asymptotic symmetry group. The main result of this section is the proof that these Hamiltonian surface charges obey an adjusted algebra which is isomorphic to the algebra of asymptotic symmetries without central extension. Throughout this section, except the existence of a conformal compactification, no boundary conditions are assumed. 

\subsection{Holographically renormalized variational principle}
\label{sec:Holographic renormalization}

The variational principle for General Relativity without matter in Al(A)dS$_4$ spacetimes is given by
\begin{equation}
S[g] =  \int_{\mathscr M} \D^4 x\, L_{\text{EH}}[g]  + \int_{{\mathscr I}} \D^3 x\, L_{\text{GHY}} + \int_{\mathscr I} \D^3 x\, L_{\text{ct}}+\int_{\mathscr I} \D^3 x\, L_\circ  .
\label{total action}
\end{equation} 
The first term in the right-hand side of \eqref{total action} is the Einstein-Hilbert action whose Lagrangian $4$-form is
\begin{equation}
\bm L_{\text{EH}}[g]  = \frac{1}{16\pi G}\left( R - 2\Lambda\right)\sqrt{-g}\, \D^4 x . \label{eq:LEH}
\end{equation}
The second term is the Gibbons-Hawking-York term
\begin{equation}
L_{\text{GHY}}[\gamma] = \frac{1}{8\pi G} \eta \sqrt{|\gamma|} K ,\qquad \eta \equiv - \text{sgn}(\Lambda). 
\end{equation} 
Here, $\mathscr M$ denotes the bulk spacetime and $\mathscr{I} = \partial \mathscr M$ its asymptotic boundary. The cosmological constant $\Lambda= \pm 3/\ell^2$ can have either sign. The extrinsic curvature $K$ of $\mathscr I$ is defined with respect to its 4-dimensional outward normal unit vector $\bm N$, $N^\mu N_\mu = \eta$. The latter is defined in the vicinity of $\mathscr I$ and determines a foliation of hypersurfaces. The induced metric on such hypersurfaces is written as $\gamma_{ab}$ and its determinant as $\gamma$. We denote as $\rho$ the Starobinsky/Fefferman-Graham coordinate that foliates the spacetime close to $\mathscr I$ (located at $\rho =0$ with $N_\mu \sim \p_\mu \rho$) \cite{Starobinsky:1982mr,Fefferman:1985aa,Skenderis:2002wp,2007arXiv0710.0919F,Papadimitriou:2010as}. Asymptotically locally (anti-)de Sitter (Al(A)dS$_4$) spacetimes are defined as spacetimes such that the boundary metric $g_{ab}^{(0)}=\rho^{2}\gamma_{ab}|_{\mathscr I} $ exists. Our notation and conventions are further explained in Appendix \ref{app:Lift}. 

The Einstein-Hilbert action is supplemented by various boundary contributions. The addition of the Gibbons-Hawking-York term is necessary and sufficient to admit a well-defined variational principle for Dirichlet boundary conditions (\textit{i.e.} $\delta \gamma_{ab}|_{\mathscr I} = \delta g^{(0)}_{ab} = 0$). The counterterm required to cancel the divergences arising from the variations of the boundary metric is given by \cite{Balasubramanian:1999re}
\begin{equation}
L_{\text{ct}}[\gamma] = \frac{1}{16\pi G}\sqrt{\frac{3}{|\Lambda|}} \sqrt{|\gamma|}\left[\frac{4\Lambda}{3}-R[\gamma]\right].
\end{equation}
Though this boundary Lagrangian does depend upon the foliation used to define the boundary metric, the surface charges do not depend upon this choice, as will be derived in Section \ref{sub:Surface charges}. This is a consequence of the absence of conformal anomaly in even-dimensional (A)dS. The choice of boundary conditions might lead to adding a finite boundary Lagrangian $L_\circ[g^{(0)}]$ which is a function of $g^{(0)}_{ab}$ and possibly other boundary fields. We assume that the associated action is diffeomorphism and Weyl invariant. As an example without additional boundary fields, one could add the gravitational Chern-Simons Lagrangian for $g^{(0)}_{ab}$. Another example will be provided in Section \ref{secflat}.

\subsection{Holographically renormalized symplectic structure}
\label{Presymplectic structure and its ambiguities}

The symplectic structure is defined in two steps. First, the presymplectic form is defined from covariant phase space methods \cite{ Lee:1990nz, Wald:1993nt, Iyer:1994ys} as we shortly review (see also \cite{Compere:2018aar , Ruzziconi:2019pzd}). Second, it is renormalized in accordance with the action \cite{Compere:2008us}. 

Taking the variation of the  Einstein-Hilbert Lagrangian \eqref{eq:LEH} we obtain 
\begin{equation}
\delta \bm{L}_{\text{EH}}[g]  =  -\frac{1}{16\pi G} (G^{\mu \nu} + \Lambda g^{\mu\nu}) \delta g_{\mu\nu} \, \D^4 x + \D \bm \Theta_{\text{EH}}[\delta g ; g] ,
\label{variation of L}
\end{equation} 
where $G_{\mu \nu} = R_{\mu\nu} - \frac{1}{2}R g_{\mu\nu}$ and $\bm\Theta_{\text{EH}}[\delta g;g]$ is the Einstein-Hilbert presymplectic potential explicitly given by
\begin{equation}
\bm \Theta_{\text{EH}} [\delta g;g] = \frac{\sqrt{-g}}{16\pi G} \left[ \nabla_\nu (\delta g)^{\mu\nu} - \nabla^\mu (\delta g)^\nu_{\phantom{\nu}\nu}\right] (\D^3 x)_{\mu} .
\label{EH presymplectic potential}
\end{equation}
We use the Grassmann even convention for the variational operator $\delta$.  The indices of $\delta g_{\mu \nu}$ are lowered and raised by $g_{\mu\nu}$ and its inverse (\textit{i.e.} $(\delta g)^{\mu\nu} = g^{\mu\alpha} g^{\nu\beta} \delta g_{\alpha\beta}$). The associated presymplectic form is defined as
\begin{equation}
\bm \omega_{\text{EH}}[\delta_1 g,\delta_2 g;g] = \delta_1 \bm\Theta_{\text{EH}}[\delta_2 g ; g] - \delta_2 \bm\Theta_{\text{EH}}[\delta_1 g ; g] .
\label{eq:presymplectic EH current}
\end{equation}

Both the Einstein-Hilbert presympletic potential and presympletic form are divergent at $\mathscr I$ if the boundary metric is varied. The renormalization of the action leads to a renormalization of the presymplectic potential/form.  This renormalization can be performed using the Starobinsky/Fefferman-Graham holographic fields $g_{ab}^{(0)}$ and $T^{ab}$ defined in Appendix \ref{Gravity in Fefferman-Graham gauge}. As proven in Appendix \ref{sec:Holographic} and in \cite{Compere:2008us}, the master equation governing this renormalization reads as  
\begin{equation}
 \Theta^{\rho}_{\text{EH}} =  \delta L_{\text{GHY}} + \delta L_{\text{ct}} - \partial_a \Theta^a_{\text{ct}}  - \frac{1}{2} \sqrt{|g^{(0)}|}T^{ab}\delta g^{(0)}_{ab} + \mathcal{O}(\rho). \label{eq:RenormalisationOfTheta}
\end{equation}
where $\Theta^a_{\text{ct}}$ is defined from
\begin{eqnarray}\label{deltaLbnd}
\delta {L}_{\text{ct}}[\gamma]  &=&  \frac{\delta L_{\text{ct}}}{\delta \gamma_{ab}} \delta \gamma_{ab} + \p_a \Theta^a_{\text{ct}}[\delta \gamma;\gamma] . 
\end{eqnarray}
We also note that
\begin{eqnarray}
0  &=& \delta {L}_\circ[g^{(0)}]  - \p_a \Theta^a_\circ[\delta g^{(0)}; g^{(0)}] -  \frac{1}{2}\sqrt{|g^{(0)}|}T^{ab}_\circ\delta g_{ab}^{(0)},\qquad T_\circ^{ab} \equiv \frac{2}{\sqrt{|g^{(0)}|}}\frac{\delta L_\circ}{\delta g^{(0)}_{ab}}.  \label{dL0}
\end{eqnarray} 
Diffeomorphism and Weyl invariance of $L_\circ$ imply
\begin{equation}
g^{(0)}_{ab} T^{ab}_{\circ} = 0, \qquad D^{(0)}_a T^{ab}_\circ=0. \label{T0}
\end{equation}

We now have the tools to define the renormalized presymplectic potential. Combining \eqref{eq:RenormalisationOfTheta} and \eqref{dL0}, we first define its component along the normal $\bm N$: 
\begin{equation}
\Theta_{\text{ren}}^\rho [\delta \phi ; \phi] \equiv  \Theta^{\rho}_{\text{EH}} - \delta L_{\text{GHY}} - \delta L_{\text{ct}} -\delta L_\circ + \partial_a \Theta^a_{\text{ct}}+\partial_a \Theta^a_\circ   \label{ren presymp pot}
\end{equation}
where $\phi$ denotes all the fields, $\phi = \{ g_{\mu\nu}, \, \gamma_{ab},\, g_{ab}^{(0)}, \dots \}$. Alternatively, $\phi$ can be taken to denote the metric $g_{\mu\nu}$ and all background structures such as the foliation. The dots refer to possible additional boundary structure used in the definition of $L_\circ$. We consider that the boundary structure is a background that is not varied. We now extend the definition to other components by geometrizing the expression as follows. We define the co-dimension 1 forms $ \bm L_{\text{GHY}}, \, \bm L_{\text{ct}}$ and $\bm L_\circ$ that admit a non-vanishing component only along the normal $\bm N$ as $ \bm L_{\text{GHY}}= L_{\text{GHY}} (\D^3x)_\rho$, $ \bm L_{\text{ct}}= L_{\text{ct}} (\D^3x)_\rho$, $ \bm L_\circ= L_\circ (\D^3x)_\rho$. Similarly, we define the co-dimension 2 forms $\boldsymbol \Theta_{\text{ct}}$ and $\boldsymbol \Theta_\circ$ with only non-vanishing components $\Theta^{\rho a}_{\text{ct}}=-\Theta^{a\rho}_{\text{ct}}=\Theta^{a}_{\text{ct}}$, $\Theta^{\rho a}_\circ =-\Theta^{a\rho}_\circ =\Theta^{a}_\circ $. We finally define the renormalized presymplectic potential as 
\begin{equation}\boxed{
\boldsymbol \Theta_{\text{ren}} [\delta \phi ; \phi] \equiv  \boldsymbol \Theta_{\text{EH}} - \delta \bm L_{\text{GHY}} - \delta \bm L_{\text{ct}} -\delta \bm L_\circ + \D \boldsymbol \Theta_{\text{ct}}+\D \boldsymbol \Theta_\circ . }\label{presc}
\end{equation}

This geometric definition fixes the two standard ambiguities that arise in the covariant phase space formalism \cite{ Lee:1990nz, Wald:1993nt, Iyer:1994ys}, namely $\bm \Theta_{\text{EH}} [\delta g ; g] \to \bm \Theta_{\text{EH}} [\delta g ; g] - \delta \bm A[\phi]+ \D \bm Y[\delta \phi ; \phi]$ where  $\bm A$ is a co-dimension 1 form and $\bm Y$ is a co-dimension 2 form that depend upon the background structure. Here $ \bm A[\phi] = \bm L_{\text{GHY}} + \bm L_{\text{ct}} + \bm L_\circ$ is fixed by the choice of boundary action that is added to the Einstein-Hilbert action in \eqref{total action} and $\bm Y[\delta \phi ; \phi]=\boldsymbol \Theta_{\text{ct}} +\boldsymbol \Theta_\circ  $ is fixed by the boundary terms arising from the variation of the boundary actions \eqref{deltaLbnd} and \eqref{dL0}. Corner terms in the variation of the action arising at $\partial \mathscr I$ are ignored here but will be discussed in Section \ref{secflat}. Under the shift \eqref{presc} the presymplectic form is renormalized as 
\begin{equation}
\bm \omega_{\text{ren}}[\delta_1 \phi,\delta_2 \phi ; \phi] \equiv \bm \omega_{\text{EH}}[\delta_1 g,\delta_2 g ; g] + \D \left( \delta_1 \bm Y [\delta_2 \phi ; \phi] - \delta_2 \bm Y [\delta_1 \phi ; \phi] \right).
\end{equation} 
The boundary term $\bm A $ does not affect the presymplectic form. The resulting pull-back on $\mathscr I$ of the renormalized presymplectic potential reads as
\begin{eqnarray}
\boxed{
\boldsymbol \Theta_{\text{ren}}[\delta \phi; \phi]\Big|_{\mathscr{I}} = - \frac{1}{2} \sqrt{|g^{(0)}|}T_{(\text{tot})}^{ab} \delta g^{(0)}_{ab} (\D^3 x)_\rho  \label{thetaren}
}
\end{eqnarray}
and the associated presymplectic form is 
\begin{eqnarray} 
\boxed{
\boldsymbol \omega_{\text{ren}}[\delta_1 \phi,\delta_2 \phi ; \phi]\Big|_{\mathscr{I}} = - \frac{1}{2} \delta_1 \left(\sqrt{|g^{(0)}|}T_{(\text{tot})}^{ab}\right)\delta_2 g^{(0)}_{ab} (\D^3 x)_\rho - (1 \leftrightarrow 2), \label{chh}
}
\end{eqnarray}  
where $T_{(\text{tot})}^{ab}=  T^{ab}+T^{ab}_\circ$ is the total stress-energy tensor.

\subsection{Canonical surface charges}
\label{sub:Surface charges}

The co-dimension 2 form $\bm k_{\xi,\text{ren}}[\delta \phi; \phi]$ associated with the vector field $\xi$ and from which one can deduce the infinitesimal surface charge difference between two neighbouring field configurations is defined as \cite{Lee:1990nz, Wald:1993nt, Iyer:1994ys,Barnich:2001jy,Barnich:2007bf,Compere:2008us} (see also the reviews \cite{Hollands:2005wt, Fischetti:2012rd,Compere:2007az,Compere:2018aar, Ruzziconi:2019pzd})
\begin{equation}
\text d \bm k_{\xi,\text{ren}}[\delta \phi; \phi] = \bm \omega_{\text{ren}}[\delta_\xi \phi,\delta \phi; \phi] ,
\label{eq:FundamentalThm}
\end{equation} 
where the metric and its variation are on-shell. The contracted variation of the metric is an infinitesimal diffeomorphism, $\delta_\xi g_{\mu\nu}= \mathcal L_\xi g_{\mu\nu}$. Remember that we are using the fields $\phi = \{ g_{\mu\nu}, \, \gamma_{ab},\, g_{ab}^{(0)}, \dots \}$ defined in Starobinsky/Fefferman-Graham gauge. A diffeomorphism preserving the gauge is parametrized by an arbitrary Weyl parameter $\sigma(x^a)$ and an arbitrary boundary diffeomorphism $\xi_{(0)}^a(x^b)$. The variation of the fields $\delta_\xi \gamma_{ab}$, $\delta_\xi g_{ab}^{(0)}$ under Weyl rescalings and boundary diffeomorphisms can be deduced from their definition, see Appendix \ref{Gravity in Fefferman-Graham gauge}. The fundamental relation \eqref{eq:FundamentalThm} defines $\bm k_{\xi,\text{ren}}[\delta \phi;\phi]$ up to an exact co-dimension $2$ form ambiguity. This ambiguity does not play any role when integrating on a $2$-sphere, assuming that the fields and the vector field are smooth, which will be the case here. 

We define a foliation $t$ (timelike for $\Lambda < 0$ and spacelike for $\Lambda > 0$) of the boundary $\mathscr I$ such that each $t$ describes a boundary $2$-sphere. The component of \eqref{eq:FundamentalThm} relevant for defining the surface charges as an integral over the boundary  $2$-sphere  $S^2_\infty$ is 
\begin{equation}
\partial_a k_{\xi,\text{ren}}^{\rho a}[\delta \phi; \phi] = \omega^\rho_{\text{ren}} [\delta_\xi \phi,\delta \phi; \phi] .
\label{fund thm rho}
\end{equation}  
Using \eqref{chh} we deduce after some algebra the explicit expression of the co-dimension 2 form
\begin{equation}
k_{\xi,\text{ren}}^{\rho a}[\delta \phi;\phi] =\delta \left( \sqrt{|g^{(0)}|} T^a_{(\text{tot}) b} \right)\xi^b_{(0)} -\frac{1}{2} \sqrt{|g^{(0)}|}\,\xi_{(0)}^a\,T^{bc}_{(\text{tot})}\delta g^{(0)}_{bc} + \mathcal{O}(\rho) .
\label{co dimension 2 AlAdS}
\end{equation} 
 The proof is given in Appendix \ref{Co-domension 2 form}. Integrating this last expression on $S^2_\infty$ leads to the surface charges of Al(A)dS$_4$ spacetimes in terms of Starobinsky/Fefferman-Graham fields, 
\begin{equation}
\boxed{
\ndelta H_\xi [\phi] = \int_{S^2_\infty} 2 (\D^2 x)_{\rho t} \left[ \delta \left( \sqrt{|g^{(0)}|} {T_{(\text{tot})b}^t} \right)\xi^b_{(0)} -\frac{1}{2} \sqrt{|g^{(0)}|}\,\xi_{(0)}^t\,T_{(\text{tot})}^{bc}\delta g^{(0)}_{bc} \right] . 
} \label{eq:FinalCharges}
\end{equation} 
 Here  $2 (\D^2 x)_{\rho t} = \frac{1}{2}\varepsilon_{\rho t AB} \D x^A \wedge \D x^B$ (see Appendix \ref{app:Lift} for our conventions on differential forms). The symbol $\ndelta$ emphasizes that the charges are not integrable. This is the first main result of this paper. While this expression appeared before in the literature in restricted contexts, it is proven in more generality here. Let us make a few comments.
\begin{itemize}[label=$\rhd$]

\item The charges \eqref{eq:FinalCharges} are finite even if the boundary metric is varied. This is a consequence of the holographic renormalization of the symplectic structure, itself deduced from the holographic renormalization of the action as explained in Sections \ref{sec:Holographic renormalization} and \ref{Presymplectic structure and its ambiguities}. 

\item The charges are generically neither conserved nor integrable in the absence of boundary conditions. These two features arise due to the presence of symplectic flux \eqref{chh} leaking through the boundary $\mathscr{I}$. If one imposes boundary conditions such that this flux vanishes (for example, Dirichlet $\delta g^{(0)}_{ab}=0$ or Neumann $\delta T_{(\text{tot})}^{ab}=0$ boundary conditions), the charges become conserved and integrable. 

\item The charges are defined for arbitrary Al(A)dS$_4$ spacetimes obeying Einstein's equations without matter and are associated with the most general vector field $\xi$ that preserves the Starobinsky/Fefferman-Graham gauge. This extends the results of \cite{Compere:2008us} (see their equation (3.15)) in three directions. 
\begin{enumerate}[label=$(\roman*)$]
\item The expression \eqref{eq:FinalCharges} is valid for arbitrary $\sigma$, which proves that diffeomorphisms inducing a Weyl transformation of the boundary metric are associated with vanishing charges, as previously announced in Section \ref{sec:Holographic renormalization}. This proves that Weyl rescalings are trivial asymptotic symmetries. We expect that this feature is specific to even dimensions while the Weyl anomaly leads to a charge associated with the Weyl frame in odd dimensions  \cite{Henningson:1998gx , Papadimitriou:2005ii}.
\item It is valid for both signs of the cosmological constant: $\Lambda> 0$ and $\Lambda < 0$.
\item It allows for field-dependent diffeomorphism generators $\xi^a_{(0)}$ and $\sigma$, which will be the case when considering the boundary conditions leading to the $\Lambda$-BMS$_4$ asymptotic symmetry algebra \cite{Compere:2019bua}, as we will discuss in Section \ref{sec:Lambda-BMS charge algebra}. 
\end{enumerate}

\item The charges were deduced from integration of \eqref{fund thm rho}. Due to the presence of background structure independent from the bulk metric, the procedure of \cite{ Lee:1990nz, Wald:1993nt, Iyer:1994ys} that brings the renormalization at the level of the co-dimension 2 form is not directly applicable here and requires a generalization which will be discussed elsewhere. 
\end{itemize}

In the presence of non-conservation and non-integrability, a prescription is required in order to define the Hamiltonian, see discussions in \cite{Wald:1999wa,Compere:2019gft}. Here, we prescribe the split of the surface charges \eqref{eq:FinalCharges} into the covariant Hamiltonian and the non-integrable surface charge (or ``heat term'') as
\begin{equation}
\ndelta H_\xi [\phi] =  \delta H_\xi[\phi] + \Xi_\xi[\delta \phi;\phi] ,
\label{split}
\end{equation} where
\begin{eqnarray}
H_\xi [\phi] &=& \int_{S_\infty^2} 2 (\D^2 x)_{\rho t}~  \left[\sqrt{|g^{(0)}|}\,g_{(0)}^{tc}T^{(\text{tot})}_{bc}\xi_{(0)}^b \right] ,\label{Ham} \\
\Xi_\xi[\delta \phi;\phi] &=& \int_{S_\infty^2} 2 (\D^2 x)_{\rho t} ~\left[-\frac{1}{2}\sqrt{|g^{(0)}|} \xi_{(0)}^t \left( T_{(\text{tot})}^{bc}\delta g_{bc}^{(0)}\right)\right] - H_{\delta\xi}[\phi]. 
\label{int and non int}
\end{eqnarray}
As we will show in the next Section \ref{sub:Charge algebra}, it is the unique definition of the Hamiltonian that leads to a charge algebra without central term. This definition also leads to the canonical conserved surface charges for all derived boundary conditions in AlAdS$_4$ spacetimes namely Dirichlet \cite{Henneaux:1985tv}, Neumann \cite{Compere:2008us} and Dirichlet-Neumann \cite{Compere:2019bua} boundary conditions.

\subsection{Charge algebra under the adjusted Dirac bracket}
\label{sub:Charge algebra}

The residual diffeomorphisms of Al(A)dS$_4$ spacetimes in Starobinsky/Fefferman-Graham gauge are the Weyl rescalings and boundary diffeomorphisms parametrized by $\sigma(x^a)$ and $\xi_{(0)}^a$, respectively. See \eqref{AKV 1} and \eqref{AKV 2} for details. Weyl rescalings are trivial asymptotic symmetries because their associated charges are zero. Instead, boundary diffeomorphisms form the algebra of non-trivial asymptotic symmetries since they are associated with finite surface charges, though these are not necessarily integrable nor conserved in the absence of restrictive boundary conditions.

From general considerations \cite{Barnich:2010eb, Barnich:2010xq , Barnich:2018gdh}, field-dependent asymptotic symmetries satisfy an algebra, called the asymptotic symmetry algebra, under the adjusted Lie bracket. More specifically, if $\xi$ and $\chi$ denote two asymptotic Killing vectors, the adjusted Lie bracket is defined as
\begin{equation}
[\xi,\chi]_\star = [\xi,\chi] - \delta_\xi \chi + \delta_\chi \xi. \label{eq:ModLieBracket}
\end{equation}
The first term in the right-hand side is the standard Lie bracket of vector fields, while the two other terms take into account the field-dependence of the asymptotic Killing vectors. 

Using the adjusted Lie bracket, we show in Appendix \ref{app:Modified Lie bracket} that any asymptotic Killing vectors $\xi$ and $\chi$ of Al(A)dS$_4$ spacetimes obey the following asymptotic symmetry algebra: 
\begin{equation}
\boxed{
\begin{split}
&[\xi(\sigma_\xi,\xi^a_{(0)}),\chi(\sigma_\chi,\chi^a_{(0)})]_\star = \hat \xi (\hat \sigma,\hat \xi_{(0)}^a) , \quad \text{with}\\
	&\left\lbrace
	\begin{split}
		&\,\,\hat \sigma = \xi_{(0)}^a\partial_a \sigma_\chi - \chi^a_{(0)}\partial_a \sigma_\xi - \delta_\xi\sigma_\chi + \delta_\chi					\sigma_\xi, \\
        &\,\,\hat \xi^a_{(0)} = \xi^b_{(0)}\partial_b \chi^a_{(0)} - \xi^b_{(0)}\partial_b \chi^a_{(0)} - \delta_\xi\chi_{(0)}^a + 						\delta_\chi\xi_{(0)}^a.
	\end{split}
	\right.
\end{split}
}
\label{eq:VectorAlgebra}
\end{equation}
This is the second main result of this paper. In particular, if the parameters of the asymptotic Killing vectors are field-independent (\textit{i.e.} $\delta_\xi\sigma_\chi = \delta_\chi\sigma_\xi = \delta_\xi\chi_{(0)}^a = \delta_\chi\xi_{(0)}^a = 0$), the asymptotic symmetry algebra reduces to the semi-direct sum $\text{Diff}(\mathscr{I}) \loplus \mathbb{R}$, where Diff($\mathscr{I}$) denotes the diffeomorphisms on the boundary $\mathscr{I}$, parametrized by $\xi^a_{(0)}$, and $\mathbb{R}$ denotes the abelian Weyl rescalings on the boundary, parametrized by $\sigma$.

Let us now discuss the charge algebra in Al(A)dS$_4$ spacetimes. Following the seminal work of \cite{Barnich:2011mi}, we define the adjusted Dirac bracket as
\begin{equation}
\lbrace H_\xi[\phi],H_\chi[\phi] \rbrace_\star \equiv \delta_\chi H_\xi[\phi] + \Xi_\chi [\delta_\xi \phi,\phi]. \label{br}
\end{equation}
In Appendix \ref{app:Charge algebra}, we show that, using this adjusted bracket, the charges of Al(A)dS$_4$ spacetime represent the algebra \eqref{eq:VectorAlgebra} without central extension, 
\begin{equation}
\boxed{
\lbrace H_\xi[\phi],H_\chi[\phi] \rbrace_\star = H_{[\xi,\chi]_\star}[\phi] . \label{eq:Algebra}
}
\end{equation} 
This is the third main result of this paper. Let us make a few comments.
\begin{itemize}[label=$\rhd$]
\item Despite the generic lack of integrability of the canonical surface charges \eqref{eq:FinalCharges}, the Hamiltonian charges \eqref{Ham} obey an algebra under the adjusted bracket \eqref{br}. It is a remarkable fact that the adjusted bracket originally defined in the context of asymptotically flat spacetimes \cite{Barnich:2011mi} applies in different situations \cite{Donnay:2016ejv , Compere:2018ylh , Donnay:2019jiz, Adami:2020amw}. As already emphasized in \cite{Barnich:2019vzx}, it would be interesting to derive this adjusted bracket from first principles in the covariant phase space formalism. It is also worth realizing this bracket in the semi-classical theory as performed in asymptotically flat spacetimes \cite{Distler:2018rwu}.

\item Let us ponder the choice of prescription for the Hamiltonian or, in other words, the split between the integrable and non-integrable parts of the surface charge that we chose in \eqref{Ham}-\eqref{int and non int}. If we define instead the Hamitonian as $H' = H + \Delta H$ and so the non-integrable heat term as $\Xi' = \Xi - \delta \Delta H$ for some $\Delta H=\Delta H_\xi [\phi]$, the charge algebra becomes
\begin{equation}
\lbrace H'_\xi[\phi],H'_\chi[\phi] \rbrace_\star = H'_{[\xi,\chi]_\star}[\phi] + K_{\xi;\chi}[\phi] , 
\label{modified algebra}
\end{equation} 
where the (non-central) extension is
\begin{equation}
K_{\xi;\chi}[\phi] \equiv  \delta_\chi \Delta H_\xi [\phi] - \delta_\xi \Delta H_\chi [\phi] - \Delta H_{[\xi, \chi]_\star}[\phi] . 
\end{equation} This extension trivially satisfies the 2-cocycle condition given by $K_{[\xi_1,\xi_2]_\star; \xi_3} + \delta_{\xi_3} K_{\xi_1;\xi_2} + \text{cyclic}(1,2,3) = 0$, which ensures that the Jacobi identity is satisfied \cite{Barnich:2011mi}. We therefore proved that the charge algebra holds for any split between a Hamiltonian and a non-integrable heat term but it generically exhibits a 2-cocycle extension. Our definition of the Hamiltonian in \eqref{Ham} is the unique definition that allows to absorb the 2-cocycle and is therefore preferred\footnote{A similar situation where the 2-cocycle can be absorbed in the redefinition of the Hamiltonian was recently discussed in \cite{Adami:2020amw} in the context of asymptotic symmetries in the vicinity of black holes horizons.}.

\item As discussed in \cite{Barnich:2011mi , Barnich:2013axa}, the charge algebra with the adjusted Dirac bracket allows to derive the flux associated with the non-conservation of the Hamiltonian charges. Indeed, taking the particular diffeomorphism generator $\chi = \partial_t$ in \eqref{eq:Algebra}, we obtain
\begin{equation}
\delta_{\partial_t} H_\xi [\phi] + \Xi_{\partial_t}[\delta_\xi \phi ; \phi] = H_{[\xi, \partial_t]_\star}[\phi] . 
\end{equation} Using $\frac{\D }{\D t}H_\xi[\phi] = \delta_{\partial_t} H_\xi [\phi] + \frac{\partial}{\partial t} H_\xi [\phi]$,  $\frac{\partial}{\partial t} H_\xi [\phi] = H_{\partial_t \xi}[\phi] = - H_{[\xi, \partial_t]_\star} [\phi]$ and the definition \eqref{thetaren}, we obtain the flux formula
\begin{equation}
\frac{\D }{\D t}H_\xi[\phi] = -\Xi_{\partial_t} [ \delta_\xi \phi ; \phi] = - \int_{S^2_\infty} i_{\p_t} \boldsymbol \Theta_{\text{ren}}[\delta_\xi \phi ; \phi] . \label{eq:FluxBalance}
\end{equation} 
This flux formula exactly reproduces Eq. (4.9) of \cite{Anninos:2010zf} and is the analogue of the flux formula discussed in \cite{Wald:1999wa} for asymptotically flat spacetimes.
\end{itemize}

\section{$\Lambda$-BMS$_{4}$ charge algebra}
\label{sec:Lambda-BMS charge algebra}

In the previous section, we obtained the surface charges of Al(A)dS$_4$ spacetimes associated with boundary diffeomorphisms and we derived their centerless asymptotic symmetry algebra. No boundary conditions were imposed. In this section, we will specialize this general result after the Dirichlet boundary gauge fixing discussed in \cite{Compere:2019bua}. The resulting asymptotic symmetry algebra will be called the $\Lambda$-BMS$_4$ algebra. Dirichlet boundary gauge can be reached from a diffeomorphism and, therefore, such gauge fixing does not restrict the dynamics. While this boundary gauge fixing could also have an application for quantum gravity in Al(A)dS$_4$, we motivate its study here mainly because it allows us to take the flat limit. As detailed in Section \ref{Flat limit of} (see also Appendix \ref{Flat limit of solution space and symmetries}), the asymptotic symmetry algebra reduces to the (generalized) BMS$_4$ algebra in the flat limit $\Lambda \to 0$ \cite{Campiglia:2014yka , Campiglia:2015yka, Compere:2018ylh,Campiglia:2020qvc}.

\subsection{Boundary gauge fixing and $\Lambda$-BMS$_{4}$ asymptotic symmetry algebra}

Al(A)dS$_4$ spacetimes admit a $3$-dimensional boundary metric $g_{ab}^{(0)}$ by definition. The $\Lambda$-BMS$_{4}$ asymptotic symmetry algebra arises after the imposition of the following Dirichlet boundary gauge fixing conditions \cite{Compere:2019bua}
\begin{equation}
g_{tt}^{(0)} = \frac{\Lambda}{3}, \qquad g_{tA}^{(0)} = 0, \qquad \sqrt{|g^{(0)}|} = \sqrt{\frac{|\Lambda|}{3}}\sqrt{\bar q}. \label{eq:LBMSConditions}
\end{equation} 
Here, we introduced a boundary structure consisting of ($i$) a boundary foliation of constant $t$ surfaces (where $t$ is timelike for AdS$_4$ and spacelike for dS$_4$) and ($ii$) a fixed measure $\sqrt{\bar q}$ ($\delta \sqrt{\bar q} = 0$) on the hypersurfaces orthogonal to the foliation. These boundary gauge fixing conditions can be reached by a diffeomorphism and do not restrict the bulk dynamics. We will denote as $\bm T$ the 3-dimensional unit vector normal to the boundary foliation, $T^a = \ell \delta^a_t$. In the following, we will restrict our analysis to the branch of solutions containing global (A)dS$_4$ where $\bar q=\mathring{q}$, the determinant of the unit sphere metric $\mathring{q}_{AB}$.

Remark that complete Dirichlet boundary conditions consist in further imposing that the remaining components of the metric, $g_{AB}^{(0)}$,  are those of the unit sphere metric \cite{Henneaux:1985tv}. We do not impose these additional (dynamically restrictive) boundary conditions here.

For definiteness, we will derive the  $\Lambda$-BMS$_4$ asymptotic symmetry algebra in Starobinsky/ Fefferman-Graham gauge though other bulk gauges could be used. The residual diffeomorphisms preserving Starobinsky/Fefferman-Graham gauge are given in \eqref{AKV 1}-\eqref{AKV 2}. Imposing the boundary gauge fixing conditions \eqref{eq:LBMSConditions} yields algebraic and differential constraints on the parameters $\sigma$ and $\xi^a_{(0)}$. Explicitly, $\mathcal{L}_{\xi} g_{tt} = 0$, $\mathcal{L}_\xi g_{tA} = 0$ and $\mathcal{L}_{\xi} \sqrt{|g^{(0)}|} =0$ lead, respectively, to
\begin{equation}
\partial_t \xi^t_{(0)} = \frac{1}{2} D_A^{(0)} \xi^A_{(0)} , \qquad \partial_t \xi^A_{(0)} = - \frac{\Lambda}{3} g^{AB}_{(0)} \partial_B \xi^t_{(0)}  , \qquad \sigma = \frac{1}{2} D_A^{(0)} \xi^A_{(0)}.
\label{eq:LBMS4constraints}
\end{equation} Using these equations, the adjusted Lie bracket \eqref{eq:ModLieBracket} gives the following commutation relations:
\begin{equation}
[\xi(\xi^t_{(0)} , \xi^A_{(0)}), \chi(\chi^t_{(0)} , \chi^A_{(0)})]_\star = \hat \xi(\hat{\xi}^t_{(0)} , \hat{\xi}^A_{(0)}) ,
\label{struc const}
\end{equation} where
\begin{equation}
\left\lbrace \,\,
\begin{split}
{\hat{\xi}}_{(0)}^t &= \xi^A_{(0)} \partial_A \chi_{(0)}^t + \frac{1}{2} \xi_{(0)}^t  D_A^{(0)} \chi^A_{(0)} - \delta_{\xi} \chi^t_{(0)} - (\xi \leftrightarrow \chi ), \\
{\hat{\xi}}_{(0)}^A &= {\xi}_{(0)}^B \partial_B \chi^A_{(0)} - \frac{\Lambda}{3} \xi_{(0)}^t g^{AB}_{(0)} \partial_B \chi_{(0)}^t  - \delta_{\xi} \chi^A_{(0)} - (\xi \leftrightarrow \chi ). \label{structure constant}
\end{split}
\right.
\end{equation} This is a direct consequence of \eqref{eq:VectorAlgebra} and \eqref{eq:LBMS4constraints}\footnote{The algebra is presented here using the adjusted Lie bracket, which is the most appropriate when the parameters are field-dependent, while it was presented under the standard Lie bracket in \cite{Compere:2019bua} (see Eqs. (4.7) and (4.8)). Note that the terms $\delta_{\xi} \chi^a_{(0)}-\delta_{\chi} \xi^a_{(0)}$ vanish in the flat limit as shown in Appendix \ref{Flat limit of solution space and symmetries}.}.

The $\Lambda$-BMS$_4$ asymptotic symmetry algebra is infinite-dimensional. Its structure constants are field-dependent due to the presence of $g^{AB}_{(0)}$. This is either called a Lie algebroid \cite{2000math.....12106F,Lyakhovich:2004kr,Barnich:2010eb,Barnich:2010xq,Barnich:2017ubf} or a soft gauge algebra \cite{Sohnius:1982rs}\footnote{In particular, the existence of the $\Lambda$-BMS$_4$ Lie algebroid is not in contradiction with recent no-go results \cite{Safari:2019zmc} that were obtained for Lie algebra deformations. Here, we have a field-dependent Lie algebroid deformation of the BMS$_4$ Lie algebra in Al(A)dS$_4$ spacetimes.}. We can give explicit expressions for the generators of the algebra in the case where $g_{AB}^{(0)}$ is the unit sphere metric $\mathring{q}_{AB}$. To do so, we use the Helmholtz decomposition for the angular part of $\xi^a_{(0)}$ as $\xi^A_{(0)} \equiv \mathring q^{AB} \partial_B \Phi (t,x^C) + \mathring \epsilon^{AB}\partial_B \Psi(t,x^C)$ where $\mathring \epsilon^{AB}$ is the Levi-Civita tensor on the unit round sphere. The solution of the $\Lambda$-BMS$_4$ constraint equations \eqref{eq:LBMS4constraints} is then given by $\Psi = \Psi(x^A)$ which spans the area-preserving diffeomorphisms on the sphere,
\begin{equation}
\Phi(t,x^A) = \left\lbrace
\begin{split}
\, & \sum_{l,m} \left[ A_{lm} \cos \left(\frac{\omega_l \,t}{\ell}\right) + \frac{1}{\ell} \, B_{lm} \sin \left(\frac{\omega_l \,t}{\ell}\right)\right] Y_{lm} (x^A)  \text{ if } \Lambda < 0,\\
\, & \sum_{l,m} \left[ A_{lm} \cosh \left(\frac{\omega_l \,t}{\ell}\right) + \frac{1}{\ell}\, B_{lm} \sinh \left(\frac{\omega_l \,t}{\ell}\right)\right] Y_{lm} (x^A)  \text{ if } \Lambda > 0.
\end{split}
\right.
\label{eq:PhiLBMS4}
\end{equation}
together with
\begin{equation}
\xi_{(0)}^t(t,x^A) = \left\lbrace
\begin{split}
\,  & \sum_{l,m} \left[ B_{lm} \cos \left(\frac{\omega_l \,t}{\ell}\right) - \ell \, A_{lm} \sin \left(\frac{\omega_l \,t}{\ell}\right)\right] \omega_l \, Y_{lm} (x^A)  \text{ if } \Lambda < 0,\\
\,  & -\sum_{l,m} \left[ B_{lm} \cosh \left(\frac{\omega_l \,t}{\ell}\right) + \ell \, A_{lm} \sinh \left(\frac{\omega_l \,t}{\ell}\right)\right] \omega_l \, Y_{lm} (x^A)  \text{ if } \Lambda > 0,
\end{split}
\right.
\label{eq:fLBMS4}
\end{equation}
where $A_{lm},B_{lm}\in\mathbb R$, for any $(l,m)\in\mathbb N\times\mathbb N$ with $|m|\leq l$, are smooth in the flat limit $\ell\to\infty$, $\omega_l^2 = \frac{1}{2}l(l+1)$ and $Y_{lm}(x^A)$ are the usual real spherical harmonics. The derivation of these formulae is outlined in Appendix \ref{app:LBMS4gen}. The subalgebra spanned by the 10 conformal Killing vectors of the boundary metric (\textit{i.e.} the $l=0$ and $l=1$ modes) is respectively $SO(1,4)$ for $\Lambda>0$ and $SO(3,2)$ for $\Lambda< 0$. In Appendix \ref{Flat limit of solution space and symmetries}, we show explictly that the flat limit of \eqref{eq:PhiLBMS4} and \eqref{eq:fLBMS4} reproduces the well-known expressions for the generators representing the (generalized) BMS$_4$ algebra $\text{Diff($S^2$)}  \loplus \text{Supertranslations}$ \cite{Campiglia:2014yka,Campiglia:2015yka,Compere:2018ylh}.

\subsection{Symplectic structure, Hamiltonians and $\Lambda$-BMS$_4$ charge algebra}
\label{Symplectic structure and charge algebra}

After imposing the boundary gauge fixing conditions \eqref{eq:LBMSConditions} and taking into account the trace conditions \eqref{T0} and \eqref{condition on energy momentum}, the pull-back at $\mathscr I$ of the renormalized presymplectic potential \eqref{presc} and presymplectic form \eqref{chh} reduce to
\begin{eqnarray}
\boldsymbol \Theta_{\text{ren}}[\delta \phi ; \phi]\Big|_{\mathscr{I}} &=&  -\frac{3\eta}{32 \pi G \ell^2}\sqrt{\bar q} \, J_{(\text{tot})}^{AB}\delta g^{(0)}_{AB} (d^3x)_\rho,\label{Thren}\\
\boldsymbol  \omega_{\text{ren}}[\delta_1 \phi,\delta_2 \phi ; \phi]\Big|_{\mathscr{I}} &=& - \frac{3\eta}{32 \pi G \ell^2}\sqrt{\bar q} \, \delta_1 J_{(\text{tot})}^{AB}\delta_2 g^{(0)}_{AB} (d^3x)_\rho - (1 \leftrightarrow 2) .\label{lambda bms form in FG}
\end{eqnarray}  
Here, we found convenient to define
\begin{equation}
\frac{3\eta }{16\pi G \ell} J^{(\text{tot})}_{AB} \equiv T_{AB}^{(\text{tot})} - \frac{1}{2} g^{(0)}_{AB} T_{C}^{(\text{tot}) C}, \qquad T^{AB}_{(\text{tot})} \equiv T^{AB}+T_{\circ}^{AB}. \label{defJtot}
\end{equation}  

Similarly, inserting the boundary gauge fixing conditions \eqref{eq:LBMSConditions} into \eqref{eq:FinalCharges}, \eqref{Ham} and \eqref{int and non int}, we obtain the $\Lambda$-BMS$_4$ surface charges 
\begin{equation}
\boxed{
\ndelta H_\xi^{\Lambda\text{-BMS}}[\phi] = \delta H_\xi^{\Lambda\text{-BMS}}[\phi] + \Xi_\xi^{\Lambda\text{-BMS}} [\delta \phi ; \phi] \label{Lambda-BMS charges 1}
}
\end{equation} where 
\begin{eqnarray}
H_\xi^{\Lambda\text{-BMS}}[\phi] &=& -\eta \ell \int_{S^2_\infty} \D^2 \Omega ~\left[  T^{(\text{tot})}_{tt}\xi^t_{(0)}  + T^{(\text{tot})}_{tB}    \xi_{(0)}^B \right]\label{Lambda-BMS charges 2}  \\
\Xi_\xi^{\Lambda\text{-BMS}} [\delta \phi ; \phi] &=& \frac{3 \eta}{32 \pi G \ell^2} \int_{S^2_\infty} \D^2 \Omega ~\left[ \xi_{(0)}^t g^{(0)}_{AB}\delta J^{AB}_{(\text{tot})} \right]  - H_{\delta\xi}^{\Lambda\text{-BMS}}[\phi] .
\end{eqnarray}
Here, $\D^2 \Omega = 2 \sqrt{\mathring q} (\D^2 x)_{\rho t}$ denotes the measure on $S^2_\infty$. This is the fourth main result of this paper. As a corollary of \eqref{eq:Algebra}, the surface charges represent the algebra of asymptotic symmetries \eqref{struc const} without central extension
\begin{equation}
\boxed{
\lbrace H^{\Lambda\text{-BMS}}_\xi[\phi], H^{\Lambda\text{-BMS}}_\chi[\phi] \rbrace_\star = H_{[\xi,\chi]_\star}^{\Lambda\text{-BMS}}[\phi]  \label{eq:AlgebraLBMS}
}
\end{equation} 
using the adjusted Dirac bracket \eqref{br}.

As discussed in more generality in Section \ref{Charges of asymptotically locally}, the non-integrability of the $\Lambda$-BMS$_4$ infinitesimal surface charges \eqref{Lambda-BMS charges 1} is caused by the presence of a non-vanishing symplectic structure  at $\mathscr{I}$ \eqref{lambda bms form in FG}. Nevertheless, the definition of the Hamiltonian \eqref{Lambda-BMS charges 2} leads to the charge algebra \eqref{eq:AlgebraLBMS} which is isomorphic to the asymptotic symmetry algebra \eqref{struc const}. The asymptotically flat limit of this charge algebra will be described in Section \ref{BMS4alg}.

\section{Flat limit of the $\Lambda$-BMS$_4$ phase space}
\label{Flat limit of}

So far we used physical quantities defined in Starobinsky/Fefferman-Graham gauge which are especially suitable to describe Al(A)dS$_4$ spacetimes but do not generically admit a well-defined flat limit $\Lambda \to 0$. It is therefore necessary to translate all results in a language appropriate to take the flat limit. Bondi gauge (see Appendix \ref{Einstein gravity in Bondi gauge}) is a convenient gauge for that purpose since we have already showed that the solution space and the asymptotic symmetries can be mapped in a bijective correspondence between Bondi and Starobinsky/Fefferman-Graham gauges \cite{Poole:2018koa,Compere:2019bua}. The asymptotic boundary $\mathscr I$ becomes the future null boundary $\mathscr I^+$ in the flat limit. In the first subsection we will review this dictionary and extend the bijective map to all dynamical quantities, namely quantities that can be constructed from the symplectic structure. We will then derive the flat limit using Bondi variables, which will require to discuss the ``corner'' terms in the variation of the action. We will show that the  symplectic structure of asymptotically locally flat space-times is obtained from the $\Lambda \to 0$ of the holographically renormalized symplectic structure once corner terms are taken into account. We will conclude with the explicit form of the (generalized) BMS$_4$ surface charge algebra.

\subsection{Dictionary between Starobinsky/Fefferman-Graham and Bondi}

A diffeomorphism between Starobinsky/Fefferman-Graham gauge \cite{Starobinsky:1982mr,Fefferman:1985aa,Skenderis:2002wp,2007arXiv0710.0919F,Papadimitriou:2010as} and Bondi gauge \cite{Bondi:1962px,Sachs:1962wk} has been explicitly constructed \cite{Poole:2018koa , Compere:2019bua}. This allows to map each dynamical quantity between the two gauges. While the flat limit $\Lambda \to 0$ of Starobinsky/Fefferman-Graham quantities such as the holographic stress-tensor is not well-defined, all quantities in Bondi gauge admit a well-defined limit. 

Using the diffeomorphism between Bondi and Starobinsky/Fefferman-Graham gauges
and imposing the boundary gauge fixing conditions \eqref{eq:LBMSConditions} or, equivalently, \eqref{LBMS COND BONDI}, the boundary metric $g_{AB}^{(0)}$ and the holographic stress-energy tensor $T_{ab}$ defined in Starobinsky/Fefferman-Graham gauge can be expressed in terms of the functions defined in the Bondi gauge as \cite{Compere:2019bua}\footnote{Remember our sign convention change for $T_{ab}$ for $\Lambda >0$ with respect to  \cite{Compere:2019bua}.} 
\begin{equation}
g_{AB}^{(0)} = q_{AB}, \qquad 
T_{ab} = \frac{3\eta}{16\pi G \ell} \left[
\begin{array}{cc}
-\frac{4}{3} M^{(\Lambda)} &- \frac{2}{3}  N^{(\Lambda)}_B \\ 
-\frac{2}{3}  N^{(\Lambda)}_A &  J_{AB} + \frac{2}{\Lambda} M^{(\Lambda)} q_{AB}
\end{array} 
\right], \label{eq:RefiningTab}
\end{equation}
where $M^{(\Lambda)} (u,x^A)$, $N^{(\Lambda)}_A (u,x^B)$ and $J_{AB} (u,x^B)$ are the boundary fields defined in \eqref{eq:hatM}-\eqref{eq:hatNA}-\eqref{eq:hatJAB} and $u$ can be substituted for $t$ at the boundary $\mathscr I$.  The conservation of the holographic stress-tensor \eqref{condition on energy momentum} is equivalent to the constraints of the Bondi mass and angular momentum \eqref{eq:EvolutionEquations} after using \eqref{eq:CAB} and the dictionary \eqref{eq:RefiningTab}. 

Noting $\frac{\partial r}{\partial x^\mu} = \delta^\rho_{\mu} + \mathcal{O}(\rho)$ where $r$ is the Bondi radial coordinate and $\rho$ the Starobinsky/ Fefferman-Graham radius \cite{Poole:2018koa , Compere:2019bua}, it is immediate that the radial component of the renormalized presymplectic potential transforms as $ \Theta_{\text{ren}}^r = \Theta_{\text{ren}}^\rho + \mathcal{O}(\rho)$. Since all dynamical quantities can be found from this presymplectic potential, it shows the dynamical equivalence of all quantities in both gauges. 

Finally note that the parameters of the $\Lambda$-BMS$_4$ generators in Starobinsky/Fefferman-Graham gauge are related to those in Bondi gauge as
\begin{equation}
\xi_{(0)}^t = f , \qquad \xi_{(0)}^A = Y^A , \qquad \sigma = \frac{1}{2} D_A Y^A .
\label{dico diffeo}
\end{equation} 
The constraints \eqref{eq:LBMS4constraints} can be expressed as 
\begin{equation}
\partial_u f = \frac{1}{2} D_A Y^A , \qquad \partial_u Y^A = - \frac{\Lambda}{3} q^{AB} \partial_B f, \qquad \omega = 0,
\label{constraints residuals}
\end{equation} 
where the Weyl transformation $\omega$ was defined in \eqref{eq:xir}.

\subsection{Flat limit of the action and corner terms}
\label{secflat}

\begin{figure}[h!]
\centering
\subfloat[AlAdS$_4$ case.]{
\begin{tikzpicture}[scale=0.55]
	\draw[white] (-7,-7) -- (-7,7) -- (7,7) -- (7,-7) -- cycle;
	\draw[thick, dashed] (0,5) -- (0,-5);
	\draw[thick, blue, densely dashed] (0,-3) ellipse (4 and 0.35);
	\draw[thick,blue] (4,3) arc (0:-180:4 and 0.35);
	\draw[thick, blue, densely dashed] (0,3) ellipse (4 and 0.35);
	\draw[thick,blue] (4,-3) arc (0:-180:4 and 0.35);
	\draw[blue] (-4, 3) node[left]{$\partial\mathscr{I}_+$};
	\draw[blue] (-4,-3) node[left]{$\partial\mathscr{I}_-$};
	\draw[thick] (4,-5) -- (4,5);
	\draw[thick] (-4,-5) -- (-4,5);
	\draw[thick, densely dashed,->] (4,5) -- (4,6);
	\draw[thick, densely dashed,->] (-4,5) -- (-4,6);
	\draw[thick, densely dashed,->] (4,-5) -- (4,-6);
	\draw[thick, densely dashed,->] (-4,-5) -- (-4,-6);
	\draw[thick] (0,5) ellipse (4 and 0.35);
	\draw[thick, densely dashed] (0,-5) ellipse (4 and 0.35);
	\draw[thick] (4,-5) arc (0:-180:4 and 0.35);
	\draw[] (4,4) node[right]{$\mathscr{I}$};
	\node[red,rotate=45] at (1.9,0.9) {$u = \text{Cst}$};
	\foreach \k in {-4.25,-4.0,...,-3.25} 
	{
	\coordinate (pk) at (\k+5.5,-\k-5);
	\coordinate (qk) at (4,-6.5-2*\k);
	\draw[red] (pk) -- (qk);
	\draw[red,-latex] (pk) -- ++(1.0,1.0);
	\draw[very thick,green!50!black,-latex] (-4,-1) -- (-5.5,-1)node[left]{$\bm N$};
	\draw[very thick,orange!90!black,-latex] (-4,-1) -- (-4,0.5)node[above]{$\bm T\qquad$};
	}
	\end{tikzpicture}
}
\hfill
\subfloat[AldS$_4$ case.]{
\begin{tikzpicture}[scale=0.55]
	\draw[white] (-7,-7) -- (-7,7) -- (7,7) -- (7,-7) -- cycle;
	\coordinate (tl) at (-5, 5);
	\coordinate (tr) at ( 5, 5);
	\coordinate (bl) at (-5,-5);
	\coordinate (br) at ( 5,-5);
	\draw[thick] (tl) -- (tr) -- (br) -- (bl) -- cycle;
	\draw[] ($(tl)!0.5!(bl)$) node[above,rotate=90]{South pole};
	\draw[] ($(tr)!0.5!(br)$) node[above,rotate=-90]{North pole};
	\draw[black!50, thick] (tr) -- (bl);
	\draw[black!50] ($(tr)!0.5!(bl)$) node[below,rotate=45,outer sep=3pt]{Cosmological horizon};
	\node[red,rotate=-45] at (-2.7,2.3) {$u = \text{Cst}$};
	\foreach \k in {-4.25,-4.0,...,-3.25} 
	{
	\coordinate (pkp) at (\k+1.5,-\k-1);
	\coordinate (qkp) at (7.5+2*\k,5);
	\coordinate (arp) at ($(pkp)+(1.0,1.0)$);
	\draw[red] (pkp) -- (qkp);
	\draw[red, -latex] (pkp) -- (arp);
	}
	\draw[] ($(tl)!0.5!(tr)$) node[above]{$\mathscr{I}$};
	\draw[thick] (tl) -- (tr) -- (br) -- (bl) -- cycle;
	\draw[] ($(tl)!0.3!(tr)$) node[circle, fill=blue, minimum size=4pt,inner sep=0pt]{};
	\draw[blue] ($(tl)!0.2!(tr)$) node[below]{$\partial\mathscr I_+$};
	\draw[] ($(tl)!0.7!(tr)$) node[circle, fill=blue, minimum size=4pt,inner sep=0pt]{};
	\draw[blue] ($(tl)!0.7!(tr)$) node[below]{$\partial\mathscr I_-$};
	\draw[very thick,green!50!black,-latex] (4.5,5) -- (4.5,6.5)node[above]{$\bm N$};
	\draw[very thick,orange!90!black,-latex] (4.5,5) -- (3,5)node[above]{$\bm T$};
\end{tikzpicture}
}
\caption{Geometry of Al(A)dS$_4$ boundaries with background structure.}
\label{fig:GeometryBndry}
\end{figure}
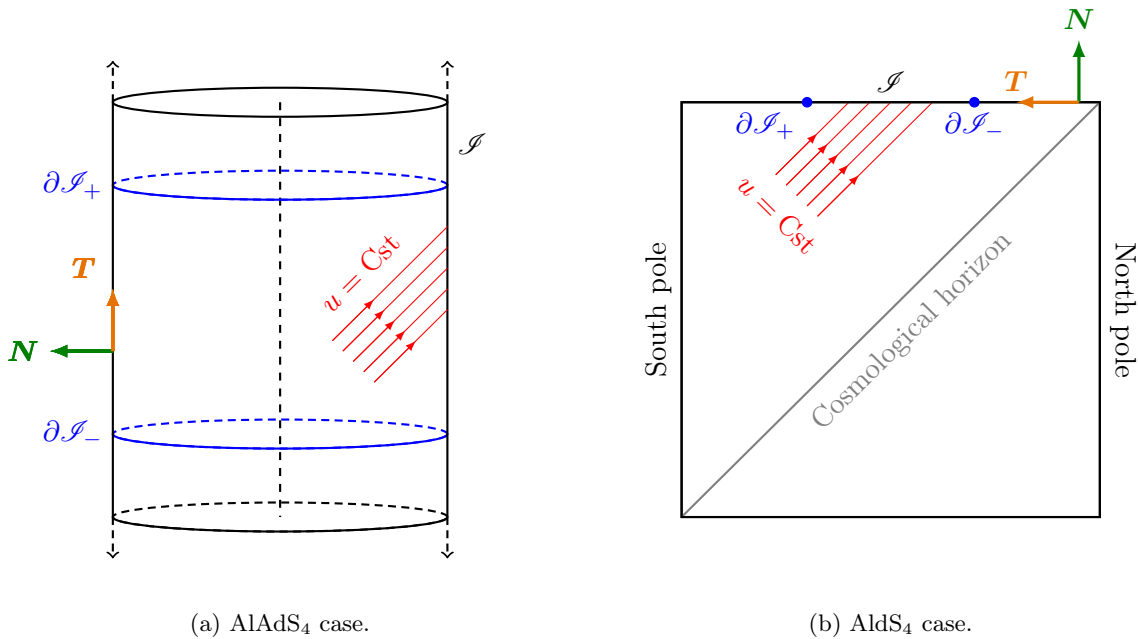

So far we have considered the variation of the action taking into account the boundary terms arising at $\mathscr I$. We have assumed the existence of a foliation $\bm T$ of the boundary with topological spheres as orthogonal surfaces of measure $\sqrt{\mathring q}$. The initial and final values of the foliation parametrized by $u$ define the  ``corner'' boundaries which we denote as $\p\!\mathscr I_+$ (at $u = u^+$) and $\p\!\mathscr I_-$ (at $u = u^-$), see Figure \ref{fig:GeometryBndry}.

Ignoring the boundary terms at fixed $u = u^\pm$, the variation of the renormalized action \eqref{total action} is given on-shell by
\begin{eqnarray}
\delta S[g] = \int_{\mathscr I} \boldsymbol \Theta_{\text{ren}}[\delta \phi ; \phi]=- \frac{3\eta}{32 \pi G \ell^2}\int_{\mathscr I} (\D^3x)_\rho \sqrt{\bar q}  \, J_{(\text{tot})}^{AB}\delta g^{(0)}_{AB} 
\end{eqnarray}  
using \eqref{Thren} obtained after boundary gauge fixing which can be expressed in either Starobinsky/ Fefferman-Graham gauge \eqref{eq:LBMSConditions} or Bondi gauge \eqref{LBMS COND BONDI}. Since such boundary gauge fixing defines the boundary structure consisting of the foliation $\bm T = \ell \p_u$ and the measure $q = \mathring{q}$ on the unit sphere, we now assume that any boundary Lagrangian $L_\circ$ in \eqref{total action} can be defined as a function $L_\circ = L_\circ[q_{AB}, \bm T , \mathring{q}]$. Using  the definitions \eqref{defJtot} and \eqref{dL0}, the dictionary \eqref{eq:RefiningTab} and the substitution $ \sqrt{\mathring{q}} (\D^3 x)_\rho = \D u \, \D^2 \Omega$ we have, equivalently, 
\begin{eqnarray}
\delta S[g] = \int_{\mathscr I} \boldsymbol \Theta_{\text{ren}}[\delta \phi ; \phi]=  \int_{\mathscr I} \D u \, \D^2 \Omega  \, \left(- \frac{\Lambda}{32 \pi G} J_{AB} +\frac{1}{\sqrt{\mathring{q}}}\frac{\delta L_\circ}{\delta q^{AB}} \right) \delta q^{AB} ,
\end{eqnarray}
where $J_{AB}$ is defined in \eqref{eq:hatJAB}.

The flat limit $\Lambda \to 0$ can be obtained as follows. One first expresses $J_{AB}$ in terms of $C_{AB}$, $q_{AB}$, $N_{AB}^{TF}$, $M$ and $N_{A}$ where $N_{AB}^{TF}$ is the trace-free part of the Bondi news tensor $N_{AB}$. In particular, all derivatives of the boundary metric $\p_u q_{AB}$ need to be expressed in terms of $C_{AB}$ using \eqref{eq:CAB}. This is the procedure used in \cite{Compere:2019bua} to obtain the asymptotically flat solution space from Al(A)dS$_4$ solutions (see Appendix \ref{app Flat limit of the solution space} for a recap). The Appendix 
\ref{Useful relations} lists useful relations for such a task. Using the form notation $\bm L_\circ = \D u \frac{\D^2 \Omega}{\sqrt{\mathring{q}}} L_\circ$, we obtain
\begin{equation}
\begin{split}
\boldsymbol \Theta_{\text{ren}}[\delta \phi ; \phi] \vert_{\mathscr I} &= \frac{\D u\, \D^2\Omega }{16\pi G} \left[ \frac{3}{2\Lambda}\partial_u N_{AB}^{TF}- D_{(A}\mathcal U_{B)} - \frac{1}{4} R[q] C_{AB} \right] \delta q^{AB}+ \frac{\delta \bm L_\circ}{\delta q^{AB}} \delta q^{AB}+ \mathcal O(\Lambda),
\end{split} \label{intermediate step flat limit}
\end{equation} 
where $\mathcal O(\Lambda)$ denotes terms that vanish in the $\Lambda \to 0$ limit. We observe that there is a pole $\sim \Lambda^{-1}$ and the flat limit is not well-defined. However, when one writes the expression in the equivalent form
\begin{eqnarray}
\boldsymbol \Theta_{\text{ren}}[\delta \phi ; \phi] \vert_{\mathscr I} &=&\frac{\D u\, \D^2\Omega }{16\pi G}  \left[ \frac{3}{2\Lambda}\partial_u (N_{AB}^{TF}\delta q^{AB}) + \frac{1}{2}\Big( N^{AB}_{TF} + \frac{1}{2} R[q]q^{AB}\Big) \delta C_{AB} - D_{(A}\mathcal U_{B)}\delta q^{AB}\right] \nonumber \\ \label{intermediate step flat limit2}
& & + \frac{\delta \bm L_\circ}{\delta q^{AB}} \delta q^{AB} + \mathcal O(\Lambda ),
\end{eqnarray} 
it becomes clear that the divergence in the variation of the action is a corner term defined at the boundaries $\p\! \mathscr I_+$, $\p \!\mathscr I_-$ of the cylinder $\mathscr I$.

In order to make these corner terms explicit, we consider the total derivative boundary Lagrangian
\begin{equation}
L_\circ= \p_u L_C[q_{AB}, \bm T, \mathring{q}] , 
\label{boundary lagrangian total derivative}
\end{equation} where the corner Lagrangian is
\begin{equation}
L_C[q_{AB},\bm T, \mathring{q}] \equiv \frac{\sqrt{\mathring{q}}}{64 \pi G} C^{AB} C_{AB} =  \frac{\sqrt{\mathring{q}}\ell^4}{64 \pi G} \p_u q_{AB} q^{AC}q^{BD}\p_u q_{CD}. \label{cornerL}
\end{equation}
The Lagrangian $L_\circ$ is of the form considered in Section \ref{sec:Holographic renormalization}. It is generally covariant on $\mathscr I$ since we can rewrite it as 
\begin{equation}
L_\circ =  \frac{\sqrt{\mathring{q}}\ell}{64 \pi G} \mathcal L_{\bm T} \left[ \mathcal L_{\bm T} g^{(0)}_{ab} P^{ac} P^{bd}\mathcal L_{\bm T}  g^{(0)}_{cd} \right]. 
\end{equation}
Here the projector onto surfaces orthogonal to $\bm T$, $P_{ab}=g_{ab}^{(0)}+ \eta T_a T_b$, is indeed a tensor on $\mathscr I$. Since $\bm L_\circ = \D \bm L_C$ where $\bm L_C = \frac{d^2 \Omega}{\sqrt{\mathring{q}}} L_C \equiv \D^2x L_C$, its stress-energy tensor \eqref{dL0} vanishes, $T_\circ^{ab} = 0$, which implies that the last term of \eqref{intermediate step flat limit2} vanishes. In particular, it obeys our hypotheses \eqref{T0}. 
 
In the action \eqref{total action}, the term associated with the Lagrangian \eqref{boundary lagrangian total derivative} can be written as
\begin{equation}
\int_{\mathscr{I}}  \D^2x \, \D u \, L_\circ = \int_{\p\! \mathscr I_+}  \D^2 x \, L_C - \int_{\p\! \mathscr I_-}  \D^2 x \,  L_C  . 
\end{equation} It is manifestly the difference of two corner terms at $\p\! \mathscr I_+$ and $\p \!\mathscr I_-$. The variation of the Lagrangian $L_\circ$ can be written as
\begin{eqnarray}
\delta L_\circ &=&  \p_u \left( \frac{\delta L_C}{\delta q_{AB}}\delta q_{AB} \right) + \p_u \Theta^C_\circ, \\
\frac{\delta L_C}{\delta q_{AB}} &=& - \frac{3 \sqrt{\mathring q} }{32\pi G \Lambda}N_{TF}^{AB},\qquad \Theta^C_\circ =  \frac{3  \sqrt{\mathring q} }{32\pi G \Lambda} \p_u \left( C^{AB}\delta q_{AB} \right). \label{divL}
\end{eqnarray}
 To obtain this result, we used \eqref{eq:CAB} and $\delta \sqrt{\mathring{q}}=0$ (see also the useful relations of Appendix \ref{Useful relations}). We now define the \emph{corner presymplectic potential} and \emph{corner presymplectic form} as
\begin{equation}
\mathbf{\Theta}^C_\circ = \D^2 x\,  \Theta^C_\circ, \qquad \boldsymbol \omega_\circ^C(\delta_1 q_{AB}, \delta_2 q_{AB}; q_{AB}) = \delta_1 \boldsymbol \Theta^C_\circ(\delta_2 q_{AB}; q_{AB}) - (1 \leftrightarrow 2).   \label{eq:DefCornerPS}
\end{equation}
We emphasize that they are defined on each sphere of the boundary $\mathscr I$, not only at $\partial \!\mathscr I_\pm$. The  presymplectic potential $\mathbf{\Theta}_\circ$ of $L_\circ$ defined in \eqref{dL0} is given by the sum $ \frac{\delta \bm L_C}{\delta q_{AB}}\delta q_{AB} + \mathbf{\Theta}^C_\circ$. 
 
We can now rewrite the divergent term in \eqref{intermediate step flat limit2} as 
\begin{eqnarray}
\frac{3\sqrt{\mathring q}}{32\pi G \Lambda} \partial_u (N_{AB}^{TF}\delta q^{AB}) &=& \p_u \left( \frac{\delta L_C}{\delta q_{AB}}\delta q_{AB} \right)  = \delta L_\circ - \p_u \Theta^C_\circ. 
\end{eqnarray} 
In parallel to the prescription \eqref{presc}, the presence of the corner terms leads us to propose the total renormalized presymplectic potential pulled-back on $\mathscr I$
\begin{equation}\boxed{
\boldsymbol  \Theta_{\text{ren,tot}}\vert_{\mathscr I} =\boldsymbol \Theta_{\text{ren}} \vert_{\mathscr I} -\delta  \D \bm L_C + \D \boldsymbol \Theta^C_\circ . \label{thetarentot}}
\end{equation} 
As already discussed under \eqref{presc}, such a prescription fixes the two standard ambiguities that arise in the covariant phase space formalism. The associated presymplectic form is
\begin{equation}
\bm \omega_{\text{ren,tot}}\vert_{\mathscr I} = \bm \omega_{\text{ren}} \vert_{\mathscr I} + \D \boldsymbol \omega^C_\circ . \label{omegatot}
\end{equation}

Before taking the flat limit, let us study how the incorporation of these corner terms impacts the $\Lambda$-BMS$_4$ charge algebra \eqref{eq:AlgebraLBMS}. The surface charge co-dimension $2$ form including the contribution of the corner presymplectic form verifies
\begin{equation}
\D \bm k_{\xi,\text{ren}}^{\text{tot}}[\delta \phi;\phi]\Big|_\mathscr{I} = \bm\omega_{\text{ren,tot}}[\delta_\xi\phi,\delta\phi;\phi]\Big|_\mathscr{I},
\end{equation}
from which we deduce
\begin{equation}
\ndelta H_{\xi,\text{tot}}^{\Lambda\text{-BMS}}[\phi] = \delta H_{\xi}^{\Lambda\text{-BMS}}[\phi] + \Xi_{\xi}^{\Lambda\text{-BMS}} [\delta \phi ; \phi] + \int_{S^2_\infty} \D^2\Omega \,\, \omega_\circ^C [\delta_\xi q_{AB},\delta q_{AB}].
\end{equation}
We dropped the dependence on $q_{AB}$ in $\omega_\circ^C$ because it involves only its variation $\delta q_{AB}$, see \eqref{divL}-\eqref{eq:DefCornerPS}. Since the corner presymplectic structure term is not integrable, we keep our definition \eqref{Lambda-BMS charges 2} of the Hamiltonian, and add the corner presymplectic structure term to the non-integrable term. Again, following the prescription of \cite{Barnich:2011mi}, the computation of the charge algebra is straightforward and gives
\begin{equation}
\begin{split}
&\lbrace H_{\xi}^{\Lambda\text{-BMS}}[\phi],H_{\chi}^{\Lambda\text{-BMS}}[\phi] \rbrace_{\star,\text{tot}} \\
&\qquad = \delta_\chi H_{\xi}^{\Lambda\text{-BMS}}[\phi] + \Xi_{\chi}^{\Lambda\text{-BMS}} [\delta_\xi \phi ; \phi] + \int_{S^2_\infty} \D^2\Omega \, \omega_\circ^C [\delta_\chi q_{AB},\delta_\xi q_{AB}] \\
&\qquad = H^{\Lambda\text{-BMS}}_{[\xi,\chi]_{\star}}[\phi] + \underbrace{\int_{S^2_\infty} \D^2\Omega \, \omega_\circ^C [\delta_\chi q_{AB},\delta_\xi q_{AB}]}_{K_{\xi;\chi}^{\Lambda\text{-BMS,tot}}[q_{AB}]}, \label{lambda bms corner}
\end{split}
\end{equation}
where $K_{\xi;\chi}^{\Lambda\text{-BMS,tot}}[q_{AB}] = -K_{\chi;\xi}^{\Lambda\text{-BMS,tot}}[q_{AB}]$ and satisfies the 2-cocycle condition $K^{\Lambda\text{-BMS,tot}}_{[\xi_1,\xi_2]_\star; \xi_3} + \delta_{\xi_3} K^{\Lambda\text{-BMS,tot}}_{\xi_1;\xi_2} \, + \, \text{cyclic}(1,2,3) = 0$. Hence the corner terms naturally give rise to a field-dependent 2-cocycle in the right-hand side of the $\Lambda$-BMS$_4$ charge algebra.

Let us now consider the flat limit of the symplectic structure. Taking $\Lambda\to 0$, the equations \eqref{thetarentot} and \eqref{omegatot} give respectively 
\begin{equation}
\boldsymbol \Theta_{\text{ren,tot}}[\delta \phi ; \phi] \vert_{\mathscr I} \,\stackrel{(\Lambda \to 0)}{=}\, \frac{\D u\, \D^2\Omega }{16\pi G}  \left[  \frac{1}{2}\Big( N^{AB}_{TF} + \frac{1}{2} R[q]q^{AB}\Big) \delta C_{AB} - D_{(A}\mathcal U_{B)}\delta q^{AB}\right]   
\end{equation}
and 
\begin{equation}
\boxed{
\bm \omega_{\text{ren,tot}}[\delta_1 \phi,\delta_2 \phi; \phi]\Big|_{\mathscr I} \stackrel{(\Lambda \to 0)}{=}  \frac{\D u \, \D^2\Omega}{16\pi G} \left[ \frac{1}{2}\delta_1 \left(N^{AB} + \frac{1}{2}R[q]q^{AB}\right) \wedge \delta_2 C_{AB} - \delta_1 \left( D_{(A}\mathcal U_{B)} \right) \wedge \delta_2 q^{AB}\right].
}\label{presymplectic current flat}
\end{equation}

This is the fifth main result of this paper. Let us make a few comments.
\begin{itemize}[label=$\rhd$]
\item Even though the flat limit of the action only requires to consider the corner action $\int_{\p \mathscr I_+} L_C - \int_{\p \mathscr I_-} L_C$ at $u=u^\pm$, the total presymplectic potential \eqref{thetarentot} and form \eqref{omegatot} are defined from the corner action at any $u$ along $\mathscr I$, which also shifts the surface charges at any $u$. 

\item The prescription \eqref{presc} defined in \cite{Compere:2008us} fixes the usual Iyer-Wald ambiguities in the definition of the presymplectic potential \cite{ Lee:1990nz, Wald:1993nt, Iyer:1994ys} using as an input the boundary counterterms defined at $\mathscr I$. Such a prescription fails to give a renormalized symplectic structure in the flat limit. Instead, we argued that the existence of a corner Lagrangian defined for each sphere on $\mathscr I$ naturally leads to the additional prescription \eqref{thetarentot} which gives a well-defined symplectic structure in the flat limit. 

\item The presymplectic form \eqref{presymplectic current flat} is the (generalized) BMS$_4$ presymplectic structure at future null infinity $\mathscr I = \mathscr{I}^+$ in asymptotically locally flat spacetimes. This expression exactly matches with the presymplectic structure obtained by regularization methods in asymptotically flat spacetimes (see equation (5.26) of \cite{Compere:2018ylh}), which is a highly non-trivial agreement. While the regularization procedure used in \cite{Compere:2018ylh} was not explicitly covariant as noted in \cite{Flanagan:2019vbl}, the procedure used here is explicitly covariant (in terms of the boundary structure) and therefore justifies a posteriori the counterterm prescription for subtracting a radial divergence used in \cite{Compere:2018ylh}. We find curious that the radially diverging term encountered in \cite{Compere:2018ylh} (see equations (5.18) and (5.19)) is structurally similar to the $\Lambda \to 0$ diverging term found here \eqref{divL}. Here, using the prescription \eqref{thetarentot}, we are able to trace the origin of this term to a corner Lagrangian \eqref{cornerL}. This Lagrangian is a kinetic action for the boundary metric of Al(A)dS$_4$ spacetimes.

\end{itemize}

\subsection{BMS$_4$ charge algebra}
\label{BMS4alg}

In the $\Lambda \mapsto 0$ limit, the $\Lambda$-BMS$_4$ algebroid reduces to the (generalized) BMS$_4$ algebra. As detailed in Appendix \ref{flatlimits}, the constraints \eqref{constraints residuals} are solved for $Y^A = Y^A(x^B)$ and $f=T(x^A) + \frac{u}{2}D_A Y^A$ and the BMS$_4$ algebra of asymptotic symmetries reads as\footnote{Note that while the leading order parameters $T,Y^A$ of the BMS$_4$ generators are field independent the subleading components of the asymptotic symmetries $\xi(T,Y)$ depend upon the fields. The appropriate bracket between the asymptotic symmetries is therefore still the adjusted bracket. We use the shorthand notations $\xi(T)\equiv \xi(T,0)$ and $\xi(Y)\equiv \xi(0,Y)$.}
\begin{eqnarray}
\left[ \xi(T_1), \xi(T_2) \right]_\star &=& 0, \nonumber \\
 \left[\xi(Y_1), \xi(T_2) \right]_\star &=& \xi(\hat T), \qquad \hat T = Y^A_1 \p_A T_2 - \frac{1}{2} D_A Y^A_1 T_2, \label{algBMS} \\
  \left[ \xi(Y_1), \xi(Y_2) \right]_\star &=& \xi(\hat Y),\qquad \hat Y^{A} = Y^B_1 \p_B Y^{A}_2 -   Y^{B}_2 \p_B Y^{A}_1. \nonumber 
\end{eqnarray}

The (generalized) BMS$_4$ charge algebra was derived in \cite{Compere:2018ylh} using a renormalization procedure in asymptotically flat spacetimes\footnote{During the course of this work, we noted two algebraic mistakes and several typos in the published version \cite{Compere:2018ylh} which will be corrected in an Erratum \cite{Compere:2020aaa}.}. Fundamentally, the infinitesimal surface charge forms $\bm k^{\text{flat}}_{\xi}$ can be defined from the flat limit of the symplectic structure $\bm \omega_{\text{ren,tot}}$ obtained in \eqref{presymplectic current flat} as
\begin{equation}
\text d \bm k^{\text{flat}}_{\xi}[\delta \phi; \phi] = \bm \omega_{\text{ren,tot}}[\delta_\xi \phi,\delta \phi; \phi] \Big|_{\Lambda = 0}.
\label{eq:FundamentalThm2}
\end{equation} 
The infinitesimal surface charge is defined after integration on the $2$-sphere $S^2_\infty$ at fixed $u$ at the asymptotic boundary $\mathscr I^+$:
\begin{equation}
\ndelta H_{\xi}^{\text{flat}} [\phi] = \int_{S^2_\infty}  \bm k^{\text{flat}}_{\xi}[\delta \phi; \phi] = \int_{S^2_\infty} 2 (\D^2 x)_{ru} \,  (k^{\text{flat}}_{\xi})^{ru}[\delta \phi; \phi] . 
\end{equation}
As already mentioned in Section \ref{sub:Surface charges}, the relevant $ru$ component of the co-dimension $2$ form can be deduced from the fundamental relation of covariant phase space formalism \eqref{eq:FundamentalThm2} by integration over $u$. We obtain the (generalized) BMS$_4$ surface charges
\begin{equation}
\boxed{
\ndelta H_{\xi}^{\text{flat}} [\phi] = \delta H_{\xi}^{\text{flat}} [\phi] + \Xi_{\xi}^{\text{flat}} [\delta \phi ; \phi] ,
}
\end{equation} where
\begin{equation} 
\begin{split}
H_{\xi}^{\text{flat}} [\phi] &= \frac{1}{16\pi G} \int_{S^2_\infty}\text{d}^2\Omega \left[ 4 f M + 2 Y^A N_A + \frac{1}{16} Y^A \partial_A (C_{BC}C^{BC}) \right],\\
\Xi_{\xi}^{\text{flat}} [\delta \phi ; \phi] &= \frac{1}{16\pi G} \int_{S^2_\infty}\text{d}^2\Omega \left[ \frac{1}{2}f \left(N^{AB}+\frac{1}{2}q^{AB} R[q] \right) \delta C_{AB}- 2\partial_{(A} f \mathcal U_{B)} \delta q^{AB}\right. \\
&\qquad \qquad \qquad \qquad \left. - f D_{(A} \mathcal U_{B)}\delta q^{AB} -\frac{1}{4}D_C D^C f C_{AB}\delta q^{AB}\right].
\end{split}
\end{equation} 
These expressions match with Eq. (5.31) of \cite{Compere:2018ylh,Compere:2020aaa}. In particular, the Hamiltonian matches with \cite{Barnich:2011mi}. 

We already derived that the (generalized) BMS$_4$ asymptotic symmetry algebra closes under the adjusted Lie bracket \eqref{eq:ModLieBracket}, see  \eqref{struc const 3} and \eqref{structure constant 3}. The BMS$_4$ charge algebra of the Hamiltonian charges $H_{\xi}^{\text{flat}} [\phi]$ is defined using the adjusted Dirac bracket \eqref{br}. After some computations, we obtain that the (generalized) BMS$_4$ charge algebra represents the (generalized) BMS$_4$ asymptotic symmetry algebra up to a (non-central) 2-cocycle, 
\begin{equation}
\boxed{
\lbrace H_{\xi}^{\text{flat}}[\phi],H_{\chi}^{\text{flat}} [\phi] \rbrace_\star = H_{[\xi,\chi]_\star}^{\text{flat}}[\phi] + K_{\xi;\chi}^\text{flat}[\phi]  }
\label{modified algebra flat case}
\end{equation} 
where
\begin{equation}
K_{\xi_1;\xi_2}^\text{flat}[\phi] = \frac{1}{16\pi G} \int_{S_\infty^2} \D^2 \Omega \, \left[ \frac{1}{2} f_1 D_A f_2 D^A R[q] + \frac{1}{2} C^{BC} f_1 D_B D_C D_D Y_2^D - ( 1 \leftrightarrow 2) \right].\label{cocycle}
\end{equation} 
The 2-cocycle condition reads as $K_{[\xi_1,\xi_2]_\star; \xi_3} + \delta_{\xi_3} K_{\xi_1;\xi_2} + \text{cyclic}(1,2,3) = 0$. We conclude with some remarks. 

\begin{itemize}[label=$\rhd$]

\item As for the Al(A)dS$_4$ charge algebra discussed in Section \ref{sub:Charge algebra}, the charge algebra \eqref{modified algebra flat case} is invariant under a simultaneous redefinition of the Hamiltonian $H^{\text{flat}}_{\xi} \mapsto H^{\text{flat}}_{\xi} + \Delta H_\xi$, of the non-integrable piece $\Xi_{\xi}^{\text{flat}} \mapsto \Xi_{\xi}^{\text{flat}} - \delta \Delta H_{\xi}$ and of the 2-cocycle by a trivial 2-cocycle,
\begin{equation}
K^\text{flat}_{\xi_1;\xi_2}[\phi] \mapsto  K^\text{flat}_{\xi_1;\xi_2}[\phi] - \delta_{\xi_1} \Delta H_{\xi_2} [\phi] + \delta_{\xi_2} \Delta H_{\xi_1} [\phi] - \Delta H_{[\xi_1, \xi_2]_\star}[\phi] . \label{trl}
\end{equation}

\item The first term in the 2-cocycle \eqref{cocycle} is non-trivial. Indeed, two supertranslations commute and therefore the central charge does not transform upon shifting the Hamiltonian with a supermomentum charge $F_{\xi[T]}$ that depends upon $q_{AB}$ and not $C_{AB}$, \textit{i.e.} $K^\text{flat}_{\xi_1(T_1);\xi_2(T_2)}[\phi] \mapsto  K^\text{flat}_{\xi_1(T_1);\xi_2(T_2)}[\phi]$ using \eqref{trl} and $\delta_T q_{AB} = 0$. Since $K^\text{flat}_{\xi_1(T_1);\xi_2(T_2)}[\phi] \neq 0$ and only depends upon $q_{AB}$ the cocycle is non-trivial. A cohomological formulation of the second term of \eqref{cocycle} can be found in \cite{Barnich:2017ubf}. After semi-classical quantization, the second term of the 2-cocyle can also be related to the non-commutativity of the double soft limit of gravitons \cite{Distler:2018rwu}.

\item The charge algebra \eqref{modified algebra flat case} cannot be straightforwardly deduced by taking the flat limit of the $\Lambda$-BMS$_4$ charge algebra \eqref{lambda bms corner} that takes the presence of corner terms into account.  Indeed, due to the subleading field-dependence of the diffeomorphism between Fefferman-Graham and Bondi gauges, the asymptotic Killing vectors do not transform as simple vectors (see \textit{e.g.} Eq. (70) of \cite{Compere:2016hzt}). This implies that the surface charge co-dimension 2 form transforms non-trivially, which leads to a shift of the objects appearing in the charge algebra that is hard to track. For example, the 2-cocycle in \eqref{lambda bms corner} does not admit a well-defined flat limit and is therefore not directly related to the 2-cocycle \eqref{cocycle} obtained in the flat case. It is simpler to take the flat limit at the level of the symplectic structure that determines all dynamical quantities rather than at the level of the charge algebra. 

\item Defining $\bar H_{\xi(T,Y)}^{\text{flat}} [\phi] \equiv H_{\xi(T,Y)}^{\text{flat}} [\phi] + \Delta H^{\text{flat}}_{\xi(T,Y)}[\phi]$ and using the shift rules explained around \eqref{trl}, one can rewrite the algebra \eqref{modified algebra flat case} with the \emph{standard} Dirac bracket as 
\begin{equation}
\lbrace \bar H_{\xi}^{\text{flat}}[\phi],\bar H_{\chi}^{\text{flat}} [\phi] \rbrace = \bar H_{[\xi,\chi]_\star}^{\text{flat}}[\phi] +R_{\xi,\chi}[\phi],
\end{equation}
where the residue
\begin{equation}
R_{\xi,\chi}[\phi] \equiv K_{\xi;\chi}^\text{flat} [\phi]  - \Delta H^{\text{flat}}_{[\xi,\chi]_\star}  [\phi] + \delta_{\chi} \Delta H^{\text{flat}}_{\xi} [\phi]  - \Xi_\chi [\delta_\xi \phi ; \phi] 
\label{Residue}
\end{equation}
is not manifestly antisymmetric. Let us now analyze this algebra at the corners $\mathscr I^+_-$ and $\mathscr I^+_+$. We impose the boundary condition $N^{AB}=N_{\text{vac}}^{AB}+o(u^{-1})$ as $u \rightarrow \pm\infty$ where $N^{AB}_{\text{vac}}$ is the vacuum contribution of the news tensor induced by super-Lorentz transformations \cite{Compere:2018ylh}, which can be built from Geroch's bidimensional Weyl tensor \cite{1977asst.conf....1G,Campiglia:2020qvc}. We also require the shear to have an asymptotically vanishing magnetic part up to the vacuum contribution \cite{Compere:2018ylh}, $C_{AB} =(u+C_\pm) N^{\text{vac}}_{AB} -2 D_A D_B C_\pm + q_{AB}D^E D_E C_\pm + o(u^{0})$ where $C_\pm$ is the supertranslation field at $\mathscr I^+_\pm$ \cite{Strominger:2013jfa}. Inspired by the shifts proposed in Eq. (5.49) of \cite{Compere:2018ylh} and Eq. (4.4) of \cite{Campiglia:2020qvc}, we prescribe
\begin{equation}
\Delta H^{\text{flat}}_{\xi(T,Y)}[\phi] \equiv \int_{S_\infty^2} \frac{\text{d}^2 \Omega}{16 \pi G} \left[ \frac{1}{2} f C_{AB} N^{AB}_{\text{vac}} + \frac{1}{2} Y^A  C_{AB} D_C C^{BC} + \frac{1}{8}Y^A\partial_A( C^B_C C^C_B) \right]. \label{DeltaHFin}
\end{equation}
The total Hamiltonian $\bar H_{\xi(T,Y)}^{\text{flat}} [\phi]$ can be written as 
\begin{equation}
\bar H_{\xi(T,Y)}^{\text{flat}} [\phi] = \int_{S_\infty^2} \frac{\text{d}^2 \Omega}{16 \pi G} \Big[ 4\, T \bar M + 2\, Y^A \bar N_A \Big],
\end{equation}
where 
\begin{align}
\bar M &= M + \frac{1}{8}  C_{AB} N_{\text{vac}}^{AB}, \\ 
\bar N_A &= N_A -u \p_A  \bar M  + \frac{1}{4} C_{AB} D_C C^{BC} + \frac{3}{32}\partial_A( C_{BC}C^{BC}). 
\end{align}
It obeys the remarkable property that $\bar H_{\xi(T,Y)}^{\text{flat}} [\phi] = 0$ for vacuum configurations where the Weyl tensor vanishes. This can be checked using the vacuum values for $M$ and $N_A$ as given in Eq. (3.26) of \cite{Compere:2018ylh}. In fact, the shift \eqref{DeltaHFin} is the unique prescription depending on $C_{AB}$, $q_{AB}$ and $N_{AB}^\text{vac}$ that obeys this property. Furthermore, inserting the shift \eqref{DeltaHFin} into \eqref{Residue}, we can show after a quite long computation that $R_{\xi,\chi}[\phi] = 0$ at the corners $\mathscr I^+_-$ and $\mathscr I^+_+$. This implies that the BMS$_4$ surface charge algebra closes under the standard Dirac bracket at $\mathscr I^+_-$ and $\mathscr I^+_+$,
\begin{equation}
\boxed{
\lbrace \bar H_{\xi}^{\text{flat}}[\phi],\bar H_{\chi}^{\text{flat}} [\phi] \rbrace \Big\vert_{\mathscr I^+_\pm} = \bar H_{[\xi,\chi]_\star}^{\text{flat}}[\phi] \Big\vert_{\mathscr I^+_\pm}.
} 
\label{alg5}
\end{equation}
This is the sixth main result of this paper. For vacuum configurations, this is trivial since all BMS$_4$ charges are zero but the algebra is valid also for non-vacuum configurations with Poincar\'e charges such as mass and angular momentum, in the presence of displacement memory and in arbitrary Lorentz and super-Lorentz frames. The result also applies at the corners of the past null boundary under similar boundary conditions. In fact, since there is no flux at spatial infinity, all BMS$_4$ generators $\bar H_{\xi}^{\text{flat}}[\phi]$ are conserved at spatial infinity. The BMS$_4$ charge algebra \eqref{alg5} is therefore realized at spatial infinity. Since the BMS$_4$ Hamiltonians are all generically non-vanishing as proven from their explicit expressions in Eq. (5.1) and Eq. (5.9) of \cite{Compere:2019gft}, we have proven that the asymptotic symmetry algebra of asymptotically flat spacetimes with non radiative boundary conditions at early and late times is the BMS$_4$ charge algebra \eqref{alg5} without central extension. This extends the result derived in \cite{Troessaert:2017jcm,Henneaux:2018cst} to include super-Lorentz asymptotic symmetries as well. 

We can also define the fluxes at $\mathscr I^+$ as
\begin{equation}
\bar F_{\xi}[\phi] = \bar H^{\text{flat}}_{\xi}[\phi]  \Big|_{\mathscr I_+^+} - \bar H^{\text{flat}}_{\xi}[\phi]  \Big\vert_{\mathscr I_-^+} = \int_{-\infty}^{+\infty} \D u\,  \p_u \bar H^{\text{flat}}_{\xi}[\phi]. 
\end{equation}
We denote the supermomenta fluxes and super-Lorentz fluxes as $\bar{P}_{T} \equiv \bar F_{\xi(T,0)}[\phi]$ and $\bar{J}_{Y} \equiv \bar F_{\xi(0,Y)}[\phi]$, respectively.  An immediate consequence of \eqref{alg5} is the following algebra of BMS$_4$ fluxes, 
\begin{equation}
\lbrace \bar{P}_{T_1},\bar{P}_{T_2} \rbrace = 0, \quad \lbrace \bar{J}_{Y_1},\bar{P}_{T_2}\rbrace = \bar{P}_{Y_1(T_2)}, \quad \lbrace \bar{J}_{Y_1},\bar{J}_{Y_2} \rbrace = \bar{J}_{[Y_1,Y_2]} 
\end{equation} 
where $Y_1(T_2)\equiv (Y_1^A\partial_A - \frac{1}{2} D_A Y^A_1)T_2$. Therefore, the algebra of BMS$_4$ fluxes represents the BMS$_4$ algebra of asymptotic symmetries \eqref{algBMS} without central extension. This is in agreement with the recent result \cite{Campiglia:2020qvc} obtained using alternative methods. 

\end{itemize}

\section{Discussion}
\label{sec:Discussion}

In the absence of boundary conditions, Al(A)dS$_4$ spacetimes admit a permeable boundary similar to the null boundary of asymptotically flat spacetimes. Arbitrary boundary diffeomorphisms are associated with finite surface Hamiltonian charges that obey a flux-balance law while Weyl rescalings admit an identically vanishing Hamiltonian charge. Diffeomorphisms breaking Starobinsky/Fefferman-Graham gauge were not considered here and might be associated to further non-trivial charges as in three-dimensional Einstein gravity \cite{Grumiller:2016pqb}. From our analysis, one can only assert that the asymptotic symmetry group of Al(A)dS$_4$ spacetimes without boundary conditions contains the group of boundary diffeomorphisms and excludes the group of Weyl rescalings. One of our main results is the proof that boundary diffeomorphism Hamiltonians represent the diffeomorphism algebra under the adjusted Dirac bracket \cite{Barnich:2011mi} without non-trivial central extension.

Dirichlet boundary gauge can be reached locally without constraining the initial value problem. It consists in fixing a boundary foliation and measure, which reduces the boundary diffeomorphism algebra to the $\Lambda$-BMS$_4$ algebroid. Dirichlet boundary gauge fixing is a weaker condition than Dirichlet boundary conditions in dimensions higher than three. In three dimensions, these boundary conditions are equivalent since the boundary metric is two-dimensional, and the ``$\Lambda$-BMS$_3$ algebra'' is therefore nothing else than the standard two copies of the Virasoro algebra (with a central extension related to the conformal anomaly) \cite{Brown:1986nw}. Instead, in four dimensions, the co-dimension two boundary metric $q_{AB}$ orthogonal to the foliation still admits two arbitrary functions (since its determinant is fixed). These two functions determine the structure constants of the $\Lambda$-BMS$_4$ algebroid.

While the flat limit of the two copies of the Virasoro algebra gives the BMS$_3$ algebra \cite{Barnich:2006av}, the flat limit of the $\Lambda$-BMS$_4$ algebroid gives the (generalized) BMS$_4$ algebra of supertranslations and super-Lorentz transformations \cite{Compere:2019bua}. Such a flat limit can be taken at the level of the solution space both in three \cite{Barnich:2012aw} and in four dimensions \cite{Compere:2019bua}. In order to promote this flat limit at the level of the action, of the symplectic structure and of the Hamiltonian charges, we found out that corner terms are necessary in addition to the standard holographic counterterms. More precisely, we introduced a co-dimension two kinetic Lagrangian for $q_{AB}$ and prescribed the boundary terms to be added to the symplectic structure, which completes the prescription of \cite{Compere:2008us}. We expect that our procedure for addressing corner boundaries in the presence of fluxes could also be useful in the context of finite boundaries, see \textit{e.g.} \cite{Freidel:2015gpa,Hopfmuller:2016scf,Donnelly:2016auv,Geiller:2017xad,Chandrasekaran:2018aop,Harlow:2019yfa,Horowitz:2019dym}.

The formulation of the (generalized) BMS$_4$ surface charge algebra requires a renormalization procedure \cite{Compere:2018ylh} which cannot be derived from a covariant prescription in terms of the bulk metric alone \cite{Flanagan:2019vbl}\footnote{Note that the necessity of such a renormalization procedure was not identified in the original work \cite{Barnich:2011mi} because super-Lorentz transformations were expanded in a meromorphic basis for which the surface charge radial divergences only appear surreptitiously at the meromorphic poles \cite{Compere:2018ylh}.}. In this paper we obtained the BMS$_4$ surface charge algebra as a contraction of the $\Lambda$-BMS$_4$ algebroid. The renormalization of the surface charges follows from holographic renormalization of the embedding Al(A)dS$_4$ spacetimes combined with a new prescription for treating corner terms in the presence of asymptotic fluxes. Such a renormalization rests on the introduction of a background structure that consists in an asymptotic bulk foliation by co-dimension one hypersurfaces, a boundary foliation by co-dimension two hypersurfaces and a boundary measure. It therefore provides a geometrical and covariant framework for the asymptotically flat renormalization procedure introduced in components in \cite{Compere:2018ylh}. 

Finally, we derived the asymptotic symmetry algebra at the corners  $\mathscr I^\pm_\pm$ of future and past null infinity of asymptotically flat spacetimes that are non-radiative at early and late times. We showed that the entire BMS$_4$ algebra including super-Lorentz transformations is realized as asymptotic symmetry algebra without non-trivial central extension. Moreover, we gave the prescription for fixing the ambiguity in the definition of the BMS$_4$ surface charges such that these charges are identically vanishing for vacuum configurations, and, such that the central extension explicitly vanishes. Conservation of the Hamiltonian charges at spatial infinity implies that the asymptotic symmetry group is realized at spatial infinity as well. We have therefore extended the derivation of the asymptotic symmetry group at spatial infinity \cite{Troessaert:2017jcm,Henneaux:2018cst} to include super-Lorentz asymptotic symmetries. As a corollary of the representation theorem, the fluxes at null infinity obey the BMS$_4$ algebra without central extension. This confirms the construction of \cite{Campiglia:2020qvc}, which can now be deduced from covariant phase space methods where renormalization is provided from the flat limit of the $\Lambda \neq 0$ holographic renormalization scheme with our new treatment of corner terms.

\section*{Acknowledgments}

We thank Francesco Alessio, Glenn Barnich, Alejandra Castro, Luca Ciambelli, Laura Donnay, Daniel Grumiller, Yannick Herfray, Yegor Kovorin, Charles Marteau, Gerben Oling, Marios Petropoulos, Aaron Poole and C\'eline Zwikel for useful discussions or comments. G.C., A.F. and R.R. are, respectively, Research Associate, Research Fellow and FRIA Research Fellow of the Fonds de la Recherche Scientifique F.R.S.-FNRS (Belgium). G.C. acknowledges support from the FNRS research credit J003620F, the IISN convention 4.4503.15 and the COST Action GWverse CA16104.

\appendix	

\section{Notations and conventions}
\label{app:Lift}


\paragraph{Geometry and integration} Throughout the text, we denote $\ell \equiv \sqrt{3/|\Lambda|}$ the Al(A)dS$_4$ radius. The Riemann tensor is defined using the conventions of \cite{Misner_1973} such that $R<0$ for AdS$_4$. By virtue of the Fefferman-Graham theorem, a conformal completion of the spacetime $(\mathscr M,g_{\mu\nu})$ exists as well as a foliation by (co-dimension 1) hypersurfaces everywhere orthogonal to a vector $\bm N$. Therefore, it exists a coordinate $\rho\in\mathbb R^+$ (timelike for AldS$_4$ spacetimes, spacelike for AlAdS$_4$ spacetimes) such that $\mathscr I_{\rho'}\equiv \lbrace \rho=\rho'\rbrace$, for any fixed $\rho'\in\mathbb R^+$, denotes an hypersurface of the foliation, and the conformal boundary $\mathscr I$ lies at $\rho=0$. When we perform integration on the manifold, we denote $\int_{\mathscr M} \text d^4 x \, (...) = \int_0^\infty \text d\rho' \int_{\mathscr I_{\rho'}} \text d^3 x\, (...)$. This convention sets the lower bound of the radial integral to be the boundary $\rho=0$. The integration measure $\D^3 x$ should be understood as a measure on the hypersurface $\mathscr I_{\rho'}$ at fixed $\rho = \rho'$. 

Furthermore, we will choose $\bm N$ as the outward unit normal vector to the foliation, meaning that $\bm N$ points from the inside of the enclosed region by $\mathscr I_{\rho'}$ to the outside, and $g_{\mu\nu}N^\mu N^\nu = \eta$ where $\eta = -\text{sgn}(\Lambda)$. We have explicitly $\bm N = N^\mu \bm \partial_\mu = - \eta \sqrt{|g_{\rho\rho}|}\bm \partial_\rho$. The second fundamental form of any $\mathscr I_{\rho'}$ is thus given by $K_{ab} = \frac{1}{2}\mathcal L_{\bm N}\gamma_{ab}$ where $\lbrace x^a\rbrace$ and $\gamma_{ab}$ denote respectively a set of coordinates and the induced metric on the hypersurface. The extrinsic curvature is finally defined as $K = \gamma^{ab}K_{ab}$.

\paragraph{Differential forms} We denote as $n\in\mathbb N_0$ the top form dimension. In the main text, it will be either the bulk spacetime dimension or the boundary spacetime dimension. We define
\begin{equation}
(\text d^{n-k}x)_{\mu_1\dots\mu_k} = \frac{1}{k!(n-k)!} \varepsilon_{\mu_1\dots\mu_k\nu_1\dots\nu_{n-k}}\text dx^{\nu_1}\wedge\dots \wedge \text dx^{\nu_{n-k}} ,
\end{equation} where $\varepsilon$ denotes the (numerically invariant) Levi-Civita symbol in $n$ dimensions. Applying the (horizontal) derivative operator $\D = \D x^\sigma \partial_\sigma$ on a co-dimension $k$ form 
\begin{equation}
\bm{A} = A^{\mu_1 \dots \mu_k} (\text d^{n-k}x)_{\mu_1\dots\mu_k} , 
\end{equation} we obtain
\begin{equation}
\D \bm{A} = \partial_\sigma A^{\mu_1 \dots \mu_{k-1}\sigma} (\text d^{n-k+1}x)_{\mu_1\dots\mu_{k-1}} .
\end{equation} 

Let us consider a four dimensional spacetime with coordinates $(\rho, x^a)$, $a=1,2,3$. A $3$-form $\bm{L}$ with respect to the instrinsic geometry of the co-dimension $1$ hypersurface $\mathscr I_{\rho_0}$ is a top-form, and can be written as $\bm{L} = L (\D^3x)$ for any $\rho_0$. Since
\begin{equation}
(\D ^3x) = \frac{1}{3!} \, \varepsilon_{abc} \, \D x^a \wedge \D x^b \wedge \D x^c =  \frac{1}{3!} \, \varepsilon_{\rho abc} \, \D x^a \wedge \D x^b \wedge \D x^c = (\D^3  x)_\rho,
\end{equation} where we set $\varepsilon_{\rho abc} \equiv \varepsilon_{abc}$, we can promote $L$ as the radial component $L^\rho \equiv  L$ of a co-dimension $1$ form with respect to the whole spacetime as $\bm{L} = L^\rho (\D^3 x)_\rho$. Similarly, since 
\begin{equation}
(\D^2 x)_a = \frac{1}{2!} \, \varepsilon_{abc} \, \D x^b \wedge \D x^c =  \frac{1}{2!} \, \varepsilon_{\rho a bc} \, \D x^b \wedge \D x^c = 2 (\D^2 x)_{\rho a} ,
\end{equation} a co-dimension $1$ form with respect to $\mathscr I_{\rho_0}$, $\bm Y = \Theta^a (\D^2 x)_a$, can be rewritten as the pull-back of a co-dimension $2$ form with respect to the spacetime, $\bm Y = Y^{\rho a} (\D^2 x)_{\rho a} +  Y^{a \rho }(\D^2 x)_{a \rho} = 2 Y^{\rho a} (\D^2 x)_{\rho a}$, where we identified $Y^{\rho a} \equiv \Theta^a $. 

\paragraph{Sign conventions} We define the variation $\delta_\xi$ of the metric as $\delta_\xi g_{\mu\nu} = +\mathcal L_{\xi} g_{\mu\nu}$, which is the opposite of the sign convention used in \cite{Barnich:2011mi}. Accordingly, the adjusted Dirac bracket is defined as \eqref{br}, which differs from Eq. (3.5) of \cite{Barnich:2011mi} by a global sign.

\section{Details of Al(A)dS$_4$ gravity}
\label{Details of Al(A)dS gravity}

\subsection{Einstein gravity in Starobinsky/Fefferman-Graham gauge}
\label{Gravity in Fefferman-Graham gauge}

Starobinsky/Fefferman-Graham gauge \cite{Starobinsky:1982mr,Fefferman:1985aa} is especially suited for the study of Al(A)dS gravity. We recapitulate a few standard results in this gauge and compute the Einstein-Hilbert presymplectic potential following the conventions of \cite{Compere:2019bua}. The line element is given by
\begin{equation}
\D s^2 =- \frac{3}{\Lambda}\frac{\D\rho^2}{\rho^2} + \gamma_{ab}(\rho,x^c) \D x^a \D x^b. \label{FG gauge}
\end{equation} 
We consider spacetimes with $\Lambda> 0$ (where $\rho$ is a timelike coordinate) or $\Lambda <0$ (where $\rho$ is a spacelike coordinate). The coordinates $(x^a) = (t,x^A)$ are defined on constant $\rho$ slices, $x^A$ representing the angular coordinates on the boundary sphere. The infinitesimal diffeomorphisms preserving  Starobinsky/Fefferman-Graham gauge are generated by vector fields $\xi^\mu$ satisfying
$\mathcal{L}_\xi g_{\rho \rho} = 0$, $\mathcal{L}_\xi g_{\rho a} = 0$. The first condition leads to $\partial_\rho \xi^\rho = \frac{1}{\rho} \xi^\rho$ which can be solved for $\xi^\rho$ as 
\begin{eqnarray}
\xi^\rho = \sigma (x^a) \rho. \label{AKV 1}
\end{eqnarray}
The second condition leads to $\rho^2 \gamma_{ab} \partial_\rho \xi^b - \frac{3}{\Lambda}\partial_a \xi^\rho =0$, 
which can be solved for $\xi^a$ as 
\begin{equation}
\xi^a = \xi_{(0)}^a (x^b) + \frac{3}{\Lambda}\partial_b \sigma \int_0^\rho \frac{\D\rho'}{\rho'} \gamma^{ab}(\rho', x^c).
\label{AKV 2}
\end{equation} 

We assume the boundary condition $\gamma_{ab}= \mathcal{O} (\rho^{-2})$, which is the statement for the spacetime to be AldS$_4$ when $\Lambda > 0$ and AlAdS$_4$ when $\Lambda < 0$. The conformal boundary lies at $\lbrace \rho=0\rbrace$, and $\lbrace \rho >0\rbrace$ is the bulk spacetime. The general asymptotic expansion that solves Einstein's equations is analytic,
\begin{equation}
\gamma_{ab} = \frac{1}{\rho^2} \Big( g_{ab}^{(0)} + \rho \, g_{ab}^{(1)} + \rho^2\, g_{ab}^{(2)} + \rho^3 \, g_{ab}^{(3)} + \mathcal{O}(\rho^4)  \Big),
\label{preliminary FG}
\end{equation} where $g_{ab}^{(i)}$ are functions of $(t, x^A)$. We call $g_{ab}^{(0)}$ the boundary metric and 
\begin{equation}
T_{ab} = \eta \frac{\sqrt{3|\Lambda|}}{16\pi G} g_{ab}^{(3)} \label{eq:StressTensor}
\end{equation} 
the holographic energy-momentum tensor \cite{deHaro:2000vlm}\footnote{Note that our sign convention differs from \cite{Compere:2019bua} for $\Lambda > 0$.}. Einstein's equations fix 
\begin{equation}
g^{(1)}_{ab} = 0,\quad g^{(2)}_{ab} = \frac{3}{\Lambda} \Big( R^{(0)}_{ab} - \frac{1}{4}R_{(0)} g^{(0)}_{ab} \Big),
\end{equation}
and all subleading terms in \eqref{preliminary FG} are determined in terms of the data $g^{(0)}_{ab}$ and $T^{ab}$ satisfying 
\begin{equation}
D_a^{(0)} T^{ab} = 0 , \quad g^{(0)}_{ab} T^{ab} =0.
\label{condition on energy momentum}
\end{equation} 
Here $D^{(0)}_a$ is the covariant derivative with respect to $g_{ab}^{(0)}$ and indices are raised with the inverse metric $g^{ab}_{(0)}$. The variation of the free data under the residual gauge transformations is given by
\begin{equation}
\delta_\xi g_{ab}^{(0)} = \mathcal{L}_{\xi^c_{(0)}} g_{ab}^{(0)} - 2 \sigma\, g_{ab}^{(0)} , \qquad
\delta_\xi T_{ab} = \mathcal{L}_{\xi^c_{(0)}} T_{ab}+ \sigma\, T_{ab}. \label{eq:TransformationTab}
\end{equation}

The  radial component of the Einstein-Hilbert presymplectic potential \eqref{EH presymplectic potential} reads in Starobin-sky/Fefferman-Graham gauge as
\begin{equation}
\begin{split}
\Theta^\rho_{\text{EH}} [\delta g;g] &=  \sqrt{\frac{3}{|\Lambda|}} \left[ -\frac{1}{\rho^3}\frac{2\Lambda}{3}\frac{\delta\sqrt{|g^{(0)}|}}{16\pi G} + \frac{1}{\rho}\left(-\frac{3}{4}\delta L_{\text{EH},(0)} + \partial_a \Theta^a_{\text{EH},(0)}\right)\right] \\
&\qquad - \frac{1}{2}\sqrt{|g^{(0)}|}\, T^{ab}\delta g_{ab}^{(0)} + \mathcal O(\rho).
\end{split}
\label{eq:Theta_rho}
\end{equation}
We denoted as $L_{\text{EH},(0)} = \frac{1}{16\pi G} \sqrt{\smash[b]{|g^{(0)}|}}R_{(0)}$ the Einstein-Hilbert Lagrangian density for the boundary metric $g_{ab}^{(0)}$ and $\Theta^a_{\text{EH},(0)}$ its associated Einstein-Hilbert presymplectic potential. 

\subsection{Einstein gravity in Bondi gauge}
\label{Einstein gravity in Bondi gauge}

Bondi gauge for Al(A)dS$_4$ spacetimes was studied in  \cite{Poole:2018koa , Compere:2019bua}. Bondi coordinates are given by $(u, r, x^A)$, where $u$ labels null hypersurfaces, $r$ is a parameter along the generating null geodesics, and $x^A$ are the transverse angular coordinates $x^A = (\theta, \phi)$. The gauge conditions are given by
\begin{equation}
g_{rA} = 0, \qquad g_{rr} = 0, \qquad \partial_r \left( \frac{\det (g_{AB})}{r^4} \right) = 0  ,
\end{equation} where $\det (g_{AB})$ denotes the determinant of the $2$-dimensional transverse metric $g_{AB}$. This last gauge condition is called the determinant condition\footnote{Notice that this determinant condition is weaker than the one used in \cite{Sachs:1962wk , 1962RSPSA.270..103S , Bondi:1962px}. Indeed, as in  \cite{Campiglia:2014yka , Campiglia:2015yka, Compere:2018ylh , Compere:2019bua}, we do not require that the leading order of $g_{AB}$ in the radial expansion is the unit sphere metric.}. The Bondi metric takes the form
\begin{equation}
\D s^2 = \frac{V}{r} e^{2\beta} \D u^2 - 2 e^{2\beta} + g_{AB} (\D x^A - U^A \D u) (\D x^B - U^B \D u)
\label{Bondi metric}
\end{equation} where $V$,  $\beta$, $U^A$ are arbitrary functions of all coordinates. The $2$-dimensional transverse metric $g_{AB}$ satisfies the determinant condition, but is otherwise arbitrary. Any metric can be written in this gauge. For example, global (A)dS$_4$ is obtained by choosing $\beta = 0$, $U^A = 0$, $V/r = (\Lambda r^2/3)-1$, $g_{AB} = r^2 \mathring q_{AB}$, where $\mathring q_{AB}$ is the unit sphere metric. Infinitesimal diffeomorphisms preserving Bondi gauge are generated by vector fields $\xi^\mu$ satisfying
\begin{equation}
\mathcal{L}_\xi g_{rr} = 0, \quad \mathcal{L}_\xi g_{rA} = 0, \quad g^{AB} \mathcal{L}_\xi g_{AB} = 4 \omega(u, x^A).
\label{eq:GaugeConstraints}
\end{equation} The last condition is a consequence of the determinant condition, and the prefactor of $4$ is introduced for convenience.  From \eqref{eq:GaugeConstraints}, we deduce
\begin{equation}
\begin{split}
\xi^u &= f, \\
\xi^A &= Y^A + I^A, \quad I^A = -\partial_B f \int_r^\infty \D r'  (e^{2 \beta} g^{AB}),\\
\xi^r &= - \frac{r}{2} (\mathcal{D}_A Y^A - 2 \omega + \mathcal{D}_A I^A - \partial_B f U^B + \frac{1}{2} f g^{-1} \partial_u g) ,\\
\end{split}
\label{eq:xir}
\end{equation} 
where $\partial_r f = 0 = \partial_r Y^A$, and $g= \det (g_{AB})$. The covariant derivative $\mathcal{D}_A$ is associated with the $2$-dimensional metric $g_{AB}$. The residual gauge transformations are parametrized by the 4 functions $\omega$, $f$ and $Y^A$ of $u$ and $x^A$.

Al(A)dS$_4$ spacetimes have $g_{AB} = \mathcal{O}(r^2)$ and they admit an analytic expansion for $g_{AB}$, namely 
\begin{equation}
g_{AB} = r^2 \, q_{AB}  + r\, C_{AB} + D_{AB} + \frac{1}{r} \, E_{AB} +\mathcal{O}(r^{-2})\label{eq:gABFallOff} ,
\end{equation} 
where each term involves a symmetric tensor whose components are functions of $(u,x^C)$. Indeed, for $\Lambda \neq 0$, the Fefferman-Graham theorem  \cite{Starobinsky:1982mr,Fefferman:1985aa,Skenderis:2002wp,2007arXiv0710.0919F,Papadimitriou:2010as} together with the map between Starobinsky/ Fefferman-Graham gauge and Bondi gauge, derived in \cite{Poole:2018koa , Compere:2019bua}, ensure that the expansion \eqref{eq:gABFallOff} leads to the most general solution to the vacuum Einstein equations. For $\Lambda = 0$, the analytic expansion \eqref{eq:gABFallOff} is an hypothesis since additional logarithmic branches might occur \cite{Winicour1985,Chrusciel:1993hx,ValienteKroon:1998vn}. This expansion does not impose any constraint on the parameters of the residual diffeomorphisms \eqref{eq:xir}. In the following, upper case Latin indices will be lowered and raised by the $2$-dimensional metric $q_{AB}$ and its inverse. The determinant condition implies $g^{AB}\p_r g_{AB}=4/r$ which imposes successively that $\det (g_{AB}) = r^2 \det (q_{AB})$, $q^{AB} C_{AB} = 0$ and
\begin{equation}
D_{AB} = \frac{1}{4} q_{AB} C^{CD} C_{CD} + \mathcal{D}_{AB} (u,x^C), \qquad
E_{AB} = \frac{1}{2} q_{AB} \mathcal{D}_{CD}C^{CD} + \mathcal{E}_{AB} (u,x^C),
\label{conseq det}
\end{equation}
with $q^{AB} \mathcal{D}_{AB} = q^{AB} \mathcal{E}_{AB} = 0$. 

The diffeomorphism between Bondi and Starobinsky/Fefferman-Graham gauges \cite{Compere:2019bua} allows to translate the boundary gauge fixing conditions \eqref{eq:LBMSConditions} into the corresponding conditions in Bondi gauge, 
\begin{equation}
\beta = o(1), \qquad U^A = o(1), \qquad \sqrt{q} = \sqrt{\bar{q}} . 
\label{LBMS COND BONDI}
\end{equation}

After boundary gauge fixing \eqref{LBMS COND BONDI}, Einstein's equations imply
\begin{equation}
\Lambda \mathcal{D}_{AB} = 0 ,
\label{equation no log}
\end{equation} as well as 
\begin{equation}
\frac{\Lambda}{3} C_{AB} = \partial_u q_{AB}, \label{eq:CAB}
\end{equation} while $\mathcal{E}_{AB}$ is unconstrained. Furthermore, the other functions of the on-shell Bondi metric \eqref{Bondi metric} are given by
\begin{equation}
\begin{split}
\beta &= \frac{1}{r^2} \left[- \frac{1}{32} C^{AB} C_{AB}   \right] + \mathcal{O}(r^{-4}), \\
U^A &= \frac{1}{r^2} \left[ - \frac{1}{2} D_B C^{AB} \right] + \frac{1}{r^3} \left[ -\frac{2}{3} N^A + \frac{1}{3} C^{AB} D^C C_{BC} \right] + \mathcal{O}(r^{-4}), \\
\frac{V}{r} &= r^2 \frac{\Lambda}{3} - \frac{1}{2} \left( R[q]  + \frac{\Lambda}{8} C_{AB} C^{AB} \right) - \frac{2M}{r} + \mathcal{O}(r^{-2}),
\label{solution space fall offs}
\end{split}
\end{equation} where $M$ and $N^A$ are functions of $(u, x^A)$ and are respectively called the Bondi mass aspect and the angular momentum aspect. In the following we will use the notation 
\begin{equation}
N_{AB} \equiv \partial_u C_{AB},\qquad \mathcal U_A \equiv -\frac{1}{2}D^B C_{AB}. 
\end{equation}
The Bondi news tensor $N_{AB} $ is symmetric and obeys $q^{AB}N_{AB} = \frac{\Lambda}{3} C^{AB}C_{AB}$. 

The Bondi mass and angular momentum satisfy the constraints
\begin{equation}
\partial_u  M^{(\Lambda)} + \frac{\Lambda}{6} D^A  N^{(\Lambda)}_A + \frac{\Lambda^2}{24} C_{AB} J^{AB} = 0 , \qquad 
\partial_u    N_A^{(\Lambda)} - \partial_A  M^{(\Lambda)} - \frac{\Lambda}{2} D^B  J_{AB} = 0 \label{eq:EvolutionEquations}
\end{equation} 
where $J_{AB}$ is traceless symmetric ($q^{AB} J_{AB} = 0$) and we defined the quantities 
\begin{align}
M^{(\Lambda)} &=  M + \frac{1}{8}{N}_{CD}C^{CD}, \label{eq:hatM} \\
N^{(\Lambda)}_A &=  N_A - \frac{3}{2\Lambda} D^B N_{AB} - \frac{3}{4} \partial_A \left(\frac{1}{\Lambda} R[q] - \frac{3}{8}  C_{CD}C^{CD}\right), \label{eq:hatNA}\\
J_{AB} &=  -\frac{3}{\Lambda^2} \Big[ \partial_u N_{AB}   -\frac{\Lambda}{2} q_{AB} C^{CD}N_{CD}  \Big] 
 +\frac{2}{\Lambda} (D_{(A}\mathcal U_{B)} - \frac{1}{2} q_{AB} D^C \mathcal U_C) \nonumber \\
&\quad +C_{AB} \Big[ \frac{1}{2\Lambda}R[q] + \frac{5}{16} C_{CD}C^{CD}  \Big] -\mathcal{E}_{AB} . \label{eq:hatJAB}
\end{align}

\subsection{Holographic renormalization procedure}
\label{sec:Holographic}

As already stated in the main text, the variational principle for Al(A)dS$_4$ Einstein gravity with Dirichlet boundary conditions ($\delta \gamma_{ab}|_{\mathscr I} = 0$) requires the incorporation of the Gibbons-Hawking-York term:
\begin{equation}
S[g] = S_{\text{EH}}[g] + S_{\text{GHY}}[\gamma], \qquad S_{\text{GHY}}[\gamma] = \frac{\eta}{8\pi G} \int_{\mathscr I} \text d^3 x \sqrt{|\gamma|} K
\end{equation}
where $\eta = -\text{sgn}(\Lambda)$ is the norm of the outward normal unit vector to $\mathscr I$, and $K$ the associated extrinsic curvature. When more general boundary conditions are imposed, the fluctuations of the boundary metric turn on radial divergences in the on-shell action. In order to study them, we introduce an infrared cut-off $\varepsilon > 0$ (which will be called \textit{regulator}). We denote by $\mathscr I_\varepsilon = \lbrace \rho = \varepsilon \rbrace$ the regulated boundary, on which will be defined all the necessary counterterms. The regulated variational principle
\begin{equation}
S_{\text{reg}}[g;\varepsilon] = \frac{1}{16\pi G}\int_{\varepsilon}^{\infty} \text d\rho' \int_{\lbrace\rho=\rho'\rbrace} \text d^3 x\, (R[g]-2\Lambda)\sqrt{|g|} + \frac{\eta}{8\pi G}  \int_{\mathscr I_\varepsilon} \text d^3 x  \sqrt{|\gamma|} K
\end{equation}
possesses on-shell two divergent pieces 
\begin{equation}
S_{\text{reg}}[g;\varepsilon] = \frac{1}{16\pi G} \sqrt{\frac{3}{|\Lambda|}} \int_{\mathscr I_\varepsilon} \text d^3 x \left[ -\frac{4\Lambda}{3}\sqrt{|g^{(0)}|}\frac{1}{\varepsilon^3} + \frac{1}{2}R_{(0)}\sqrt{|g^{(0)}|} \frac{1}{\varepsilon} + \mathcal{O}(\varepsilon) \right] \label{eq:Sreg}
\end{equation}
when $\varepsilon$ tends to zero. The holographic renormalization procedure amounts to supply the regulated variation principle by a counterterm action $S_{\text{ct}}[\gamma; \varepsilon] = \int_{\mathscr I_\varepsilon} \D^3 x\, L_{\text{ct}}[\gamma]$. The counter-term Lagrangian $L_{\text{ct}}$ is required to be a top-form with respect to the regulated hypersurface $\mathscr I_\varepsilon$. The latter is built up from covariant objects defined on $\mathscr I_\varepsilon$, but is not required to be covariant with respect to the bulk geometry. In particular, it will involve the metric $\gamma_{ab}(\varepsilon,x^c)$ only. The renormalization requirement imposes that $S_{\text{reg}}[g;\varepsilon] + S_{\text{ct}}[\gamma;\varepsilon] = \mathcal O(\varepsilon^0)$ on-shell, after expanding in power series of $\varepsilon$. The working counter-term has been prescribed in \cite{Balasubramanian:1999re} and is given by
\begin{equation}
S_{\text{ct}}[\gamma;\varepsilon] = \int_{\mathscr I_\varepsilon} \text d^3 x\, L_{\text{ct}}[\gamma],\quad L_{\text{ct}}[\gamma] = \frac{1}{16\pi G}\sqrt{\frac{3}{|\Lambda|}} \left[\frac{4\Lambda}{3}\sqrt{|\gamma|}-R[\gamma]\sqrt{|\gamma|}\right].
\end{equation}
Evaluating $L_{\text{ct}}$ on-shell and expanding in $\varepsilon$ yields
\begin{equation}
L_{\text{ct}} = \frac{1}{16\pi G} \sqrt{\frac{3}{|\Lambda|}} \left[ \frac{4\Lambda}{3}\sqrt{|g^{(0)}|}\frac{1}{\varepsilon^3} - \frac{1}{2}R_{(0)}\sqrt{|g^{(0)}|} \frac{1}{\varepsilon} + \mathcal{O}(\varepsilon) \right]. \label{eq:LctInSeries}
\end{equation}
Hence $S_{\text{ren}}[g;\varepsilon] + S_{\text{ct}}[\gamma;\varepsilon] = \mathcal O(\varepsilon)$. Remark that when the regulator is sent to zero, the full action is zero on-shell. We define the presymplectic potential ${\Theta}^a_{\text{ct}} [\delta\gamma;\gamma]$ associated with $L_{\text{ct}}$ through $\delta {L}_{\text{ct}} = \frac{\delta {L}_{\text{ct}}}{\delta \gamma^{ab}} \delta \gamma^{ab} + \partial_a {\Theta}^a_{\text{ct}} [\delta\gamma;\gamma]$, where $\frac{\delta {L}_{\text{ct}}}{\delta \gamma^{ab}} $ is the Euler-Lagrange derivative of $L_{\text{ct}}$ with respect to $\gamma^{ab}$. It is given explicitly by 
\begin{equation}
\begin{split}
{\Theta}^a_{\text{ct}}[\delta \gamma ; \gamma]\Big|_{\mathscr I_\varepsilon} &= -\frac{1}{16 \pi G} \sqrt{\frac{3}{|\Lambda|}} \sqrt{|\gamma|} \left[D_b (\delta \gamma )^{ab} - \gamma^{ab} D_b {(\delta \gamma)^c}_c \right]\Big|_{\mathscr I_\varepsilon} \\
&= -\frac{1}{16 \pi G} \varepsilon \sqrt{|g|} \left[D_b (\delta \gamma )^{ab} - \gamma^{ab} D_b {(\delta \gamma)^c}_c \right]\Big|_{\mathscr I_\varepsilon} \\
&= - \rho \, \Theta^a_{\text{EH}} [\delta \gamma; \gamma]\Big|_{\mathscr I_\varepsilon}
\end{split}
\end{equation} where $D_a$ is the Levi-Civita connection with respect to $\gamma_{ab}$. Therefore, 
\begin{equation}
{\Theta}^a_{\text{ct}}[ \delta \gamma ; \gamma]\Big|_{\mathscr I_\varepsilon} = -\sqrt{\frac{3}{|\Lambda|}} \Theta^a_{\text{EH},(0)}[\delta g^{(0)};g^{(0)}]\frac{1}{\varepsilon} + \mathcal O(\varepsilon). 
\end{equation}
Let us now show that this holographic renormalization process implies a renormalization of the presymplectic potential involving the ambiguities of the covariant phase space methods discussed in Section \ref{Presymplectic structure and its ambiguities}. On-shell, we have
\begin{equation}
\begin{split}
\delta S_{\text{EH}}^{\text{reg}} &= \int_{\rho\geq \varepsilon} \text d^4 x \, \partial_\mu \Theta_{\text{EH}}^\mu [\delta g ; g] = -\int_{\mathscr I_\varepsilon} \text d^3 x \, \Theta^\rho_{\text{EH}}[\delta g ; g] , \label{eq:DeltaSEHreg}
\end{split}
\end{equation}
where the minus sign in the last equality is due to the fact that we integrate on $\rho$ from the boundary to the bulk, which gives the negative orientation to the Stokes formula. The resulting integrand is only the $\rho$ component of $\Theta_{\text{EH}}$ since the outward normal to the regulating surface is collinear to $\bm \partial_\rho$. Evaluating \eqref{eq:Theta_rho} on $\mathscr I_\varepsilon$ and taking \eqref{eq:LctInSeries} into account, we can show by direct calculation that the renormalized presymplectic potential defined as
\begin{equation}
\delta (S_{\text{reg}}+S_{\text{ct}})\Big|_{\mathscr I_\varepsilon} = -\int_{\mathscr I_\varepsilon} \text d^3 x \, \Theta^\rho_{\text{ren}}[\delta \phi;\phi;\varepsilon]
\end{equation}
satisfies the following identity
\begin{equation}
\Theta_{\text{ren}}^\rho[\delta \phi; \phi;\varepsilon] = \Big[\Theta^{\rho}_{\text{EH}} - \delta L_{\text{GHY}} - \delta L_{\text{ct}} + \partial_a \Theta^a_{\text{ct}}  \Big]\Big|_{\mathscr I_\varepsilon} = \Big[- \frac{1}{2} \sqrt{|g^{(0)}|}T^{ab}\delta g^{(0)}_{ab}\Big]\Big|_{\mathscr I_\varepsilon} + \mathcal{O}(\varepsilon),
\end{equation}
which finally demonstrates \eqref{eq:RenormalisationOfTheta}.

\subsection{Derivation of the surface charges}
\label{Co-domension 2 form}

In this Appendix, we show that the co-dimension 2 form \eqref{co dimension 2 AlAdS} satisfies the fundamental relation \eqref{fund thm rho} at $\mathscr{I} \equiv \{ \rho = 0 \}$. We start by computing the right-hand side of \eqref{fund thm rho}. From \eqref{eq:TransformationTab}, we work out $\delta_\xi \sqrt{|g^{(0)}|}$ and $\delta_\xi T^{ab}$ as 
\begin{align}
\delta_\xi \sqrt{|g^{(0)}|} &= \frac{1}{2}\sqrt{|g^{(0)}|} \, g^{ab}_{(0)}\delta_\xi g_{ab}^{(0)} = \sqrt{|g^{(0)}|} ( D_a^{(0)}\xi_{(0)}^a - 3 \sigma_\xi ), \label{eq:deltaG0} \\
\delta_\xi T^{ab} &= \delta_\xi (g^{ac}_{(0)} T_{cd} g^{db}_{(0)}) = \mathcal L_{\xi_{(0)}^c} T^{ab} + 5 \sigma_\xi T^{ab}. \label{eq:deltaTab}
\end{align} 
We recall that $T_{ab}$ is subjected to the constraints \eqref{condition on energy momentum}.
We also need $\delta_\xi T_{ab}^\circ = \delta_{\xi^c_{(0)}} T_{ab}^\circ + \delta_{\sigma_\xi}T_{ab}^\circ$. Since $L_\circ$ is invariant under diffeomorphisms, $\delta_{\xi^c_{(0)}} T_{ab}^\circ = \mathcal L_{\xi^c_{(0)}} T_{ab}^\circ$.
Requiring further that the conditions \eqref{T0} are preserved also leads to $\delta_{\sigma_\xi}T_{ab}^\circ =  \sigma_\xi T_{ab}^\circ$. Hence $T_{ab}^\circ$ shares the same variation as $T_{ab}$, and furthermore,
\begin{equation}
\delta_\xi T_\circ^{ab} =  \mathcal L_{\xi_{(0)}^c} T_{\circ}^{ab} + 5 \sigma_\xi T_\circ^{ab}. \label{eq:deltaTab0}
\end{equation}
Defining the shorthand notation $T^{ab}_{(\text{tot})} \equiv T^{ab}+ T^{ab}_\circ$ as in the main text, we get 
\begin{align}
\delta \Theta_{\text{ren}}[\delta_\xi \phi; \phi] &= - \delta\left(\sqrt{|g^{(0)}|} T^{ab}_{(\text{tot})}\right)D_a^{(0)}\xi^{(0)}_b -  \sqrt{|g^{(0)}|} T^{ab}_{(\text{tot})} \delta \left(D_a^{(0)}\xi^{(0)}_b\right) + \mathcal{O}(\rho),\\
-\delta_\xi \Theta_{\text{ren}}[\delta \phi; \phi] &=\frac{1}{2} \sqrt{|g^{(0)}|} \left( D^{(0)}_c \xi^c_{(0)} T^{ab}_{(\text{tot})} + \mathcal L_{\xi_{(0)}} T^{ab}_{(\text{tot})}\right)\delta g_{ab}^{(0)} +  \sqrt{|g^{(0)}|} T^{ab}_{(\text{tot})}\delta \left(D_a^{(0)}\xi^{(0)}_b\right) + \mathcal{O}(\rho). \nonumber
\end{align}
The left-hand side reads as
\begin{equation}
\begin{split}
\partial_a k^{\rho a}_{\xi,\text{ren}}[\delta \phi;\phi] &= - \delta \left(\sqrt{|g^{(0)}|}T^{ab}_{(\text{tot})}\right)D_a^{(0)} \xi^{(0)}_b - \sqrt{|g^{(0)}|} T^{ab}_{(\text{tot})} \delta g_{bc}^{(0)} D_a^{(0)}\xi^c_{(0)} \\
&\phantom{=\,\,} + \frac{1}{2} \sqrt{|g^{(0)}|} D_a^{(0)}\xi^a_{(0)} T^{bc}_{(\text{tot})} \delta g_{bc}^{(0)} + \frac{1}{2} \sqrt{|g^{(0)}|} \xi^a_{(0)} D_a^{(0)} T^{bc}_{(\text{tot})}\delta g_{bc}^{(0)} +\mathcal{O}(\rho).
\end{split}
\end{equation}
Since $\mathcal L_{\xi_{(0)}^c} (T^{ab}_{(\text{tot})}) = \xi_{(0)}^c D^{(0)}_c T^{ab}_{(\text{tot})} - 2 T^{c(a}_{(\text{tot})}D_c^{(0)} \xi_{(0)}^{b)}$, we have
\begin{equation}
\begin{split}
&\partial_a k^{\rho a}_{\xi,\text{ren}}[\delta \phi ; \phi] - \omega^\rho_{\text{ren}} [\delta_\xi \phi,\delta\phi;\phi]\\
&\quad = \frac{1}{2}\sqrt{|g^{(0)}|} \left(\mathcal L_{\xi_{(0)}^c} T^{ab}_{(\text{tot})}\delta g_{ab}^{(0)} + 2 T^{ab}_{(\text{tot})}\delta g_{bc}^{(0)}D_a^{(0)} \xi^c_{(0)} - \xi^c_{(0)} D_c^{(0)}T^{ab}_{(\text{tot})} \delta g_{ab}^{(0)}\right) + \mathcal{O}(\rho)\\
&\quad = \mathcal{O}(\rho),
\end{split}
\end{equation}
which proves the result.

\subsection{Adjusted Lie bracket}
\label{app:Modified Lie bracket}

In this Appendix, we demonstrate the relations \eqref{eq:VectorAlgebra} based on the adjusted Lie bracket \eqref{eq:ModLieBracket}. Let us denote by $\xi[g]$ and $\chi[g]$ two asymptotic Killing vectors of the form \eqref{AKV 1} and \eqref{AKV 2}. By hypothesis, if $\xi$ and $\chi$ are preserving the Starobinsky/Fefferman-Graham gauge, they satisfy
\begin{equation}
\left\lbrace
\begin{split}
\,\, \xi^\rho &= \rho \sigma_\xi(x^a), \quad \partial_\rho \xi^a = \frac{3}{\Lambda}\frac{1}{\rho}\gamma^{ab}\partial_b \sigma_\xi, \quad \lim_{\rho\to 0} \xi^a = \xi^a_{(0)}(x^b), \\
\,\, \chi^\rho &= \rho \sigma_\xi(x^a), \quad \partial_\rho \xi^a = \frac{3}{\Lambda}\frac{1}{\rho}\gamma^{ab}\partial_b \sigma_\chi, \quad \lim_{\rho\to 0} \chi^a = \chi^a_{(0)}(x^b). \\
\end{split} \right. \label{eq:RefVec}
\end{equation}
As a result, the computation of $[\xi,\chi]^\rho_\star$ is straightforward and gives
\begin{equation}
\frac{1}{\rho}[\xi,\chi]^\rho_\star = \left( \xi^a\partial_a\sigma_\chi - \chi^a\partial_a\sigma_\xi\right) - \delta_\xi\sigma_\chi + \delta_\chi\sigma_\xi.
\end{equation}
Taking a derivative with respect to $\rho$, and using $\mathcal L_{[\xi,\chi]_\star}g_{\rho a}=0$, we get
\begin{equation}
\partial_\rho \left( \frac{1}{\rho}[\xi,\chi]^\rho_\star \right) = \partial_\rho \xi^a \partial_a \sigma_\chi -\partial_\rho \chi^a\partial_a\sigma_\xi = 0,
\end{equation}
which shows that $[\xi,\chi]^\rho_\star = \rho \hat \sigma$, and
\begin{equation}
\hat \sigma = \frac{1}{\rho}[\xi,\chi]^\rho_\star \Big|_{\rho=0} =  \xi_{(0)}^a\partial_a \sigma_\chi - \chi^a_{(0)}\partial_a \sigma_\xi - \delta_\xi\sigma_\chi + \delta_\chi\sigma_\xi.
\end{equation}
Let us now consider the transverse components. By evaluating the commutator at leading order in $\rho$, we derive that
\begin{equation}
\hat \xi_{(0)}^a = \lim_{\rho\to 0} [\xi,\chi]^a_\star = [\xi_{(0)},\chi_{(0)}]^a - \delta_\xi\chi_{(0)}^a + \delta_\chi\xi_{(0)}^a.
\end{equation}
Recalling that $\delta_\xi \gamma^{ab} = \mathcal L_\xi \gamma^{ab} = \rho\sigma_\xi \partial_\rho \gamma^{ab} + \xi^c \partial_c \gamma^{ab} - 2 \gamma^{c(a}\partial_c \xi^{b)}$ and using explictly \eqref{eq:RefVec} to express $\partial_\rho \xi^a$ and $\partial_\rho \chi^b$ in terms of $\sigma_\xi$ and $\sigma_\chi$ respectively, a direct computation yields
\begin{equation}
\partial_\rho \left( [\xi,\chi]^a_\star \right) = \frac{3}{\Lambda}\frac{1}{\rho}\gamma^{ab} \partial_b \hat \sigma.
\end{equation} Putting everything together, this demonstrates \eqref{eq:VectorAlgebra}. 

\subsection{Charge algebra}
\label{app:Charge algebra}

In this Appendix, we prove the charge algebra \eqref{eq:Algebra} of Al(A)dS$_4$ spacetimes. The computation is on-shell so, in particular, the relations \eqref{condition on energy momentum} hold. Let us start by working out $\delta_\chi H_\xi[\phi]$. The computation is direct using \eqref{eq:deltaG0} and \eqref{eq:deltaTab}:
\begin{equation}
\begin{split}
\delta_\chi H_\xi[\phi] &=  \int_{S_\infty^2} 2 (\D^2 x)_{\rho t}~ \Big[  \sqrt{|g^{(0)}|} D_d^{(0)}(\chi^d_{(0)} T^{tb}_{(\text{tot})}) g^{(0)}_{bc}\xi^c_{(0)} - \sqrt{|g^{(0)}|} T^{bd}_{(\text{tot})}D_d^{(0)}\chi^t_{(0)} g_{bc}^{(0)} \xi^c_{(0)} \\
&\quad +  \sqrt{|g^{(0)}|} T^{tb}_{(\text{tot})} (D_c^{(0)}\chi_b^{(0)}) \xi^c_{(0)} \Big] + H_{\delta_\chi\xi}[\phi] .
\end{split}
\end{equation}
Obtaining the second term is just a matter of replacement
\begin{equation}
\Xi_\chi [\delta_\xi \phi;\phi] = \int_{S_\infty^2} 2 (\D^2 x)_{\rho t}~ \Big[  - \sqrt{|g^{(0)}|} \chi^t_{(0)} T^{bc}_{(\text{tot})} D_b^{(0)}\xi_c^{(0)} \Big] - H_{\delta_\xi\chi}[\phi] .
\end{equation}
Summing both contributions and using the fact that $T^{ab}_{(\text{tot})}$ is divergence-free, we get
\begin{align}
&\delta_\chi H_\xi[\phi] + \Xi_\chi[\delta_\xi \phi;\phi] \nonumber \\
& =  H_{\delta_\chi\xi}[\phi] - H_{\delta_\xi\chi}[\phi] \nonumber \\
&\qquad + \int_{S_\infty^2} 2 (\D^2 x)_{\rho t}~  \left[\sqrt{|g^{(0)}|} \, T^t_{(\text{tot})b} \left(\xi^c_{(0)} D_c^{(0)}\chi^b_{(0)} - \chi^c_{(0)} D_c^{(0)}\xi_{(0)}^b\right) - 2 \partial_b \left( \sqrt{|g^{(0)}|}\chi^{[t}_{(0)} T^{b]}_{(\text{tot})c}\xi^c_{(0)} \right) \right] \nonumber \\
& =  H_{\delta_\chi\xi}[\phi] - H_{\delta_\xi\chi}[\phi] + \int_{S_\infty^2} 2 (\D^2 x)_{\rho t}~ \left[  \sqrt{|g^{(0)}|} \,  T^t_{(\text{tot})b} [\xi_{(0)},\chi_{(0)}]^b  - 2 \partial_b \left( \sqrt{|g^{(0)}|}\chi^{[t}_{(0)} T^{b]}_{(\text{tot})c} \xi^c_{(0)} \right) \right] \nonumber \\
& = H_{[\xi,\chi]}[\phi] + H_{\delta_\chi\xi}[\phi] - H_{\delta_\xi\chi}[\phi]+ \int_{S_\infty^2} 2 (\D^2 x)_{\rho t}~\left[ - 2 \partial_b \left( \sqrt{|g^{(0)}|}\chi^{[t}_{(0)} T^{b]}_{(\text{tot})c}\xi^c_{(0)} \right) \right] \nonumber \\
& = H_{[\xi,\chi]_\star}[\phi] + \int_{S_\infty^2} 2 (\D^2 x)_{\rho t}~\left[ - 2\partial_B \left( \sqrt{|g^{(0)}|}\chi^{[t}_{(0)} T^{B]}_{(\text{tot})c}\xi^c_{(0)} \right) \right].
\end{align}
The last term is a total derivative on $S^2_\infty$ and can be discarded. This proves the result.

\section{$\Lambda$-BMS$_4$ generators around on the unit round sphere}
\label{app:LBMS4gen}
In this Appendix, we show how to integrate the constraint equations \eqref{eq:LBMS4constraints} giving rise to the $\Lambda$-BMS$_4$ generators when the background boundary metric $g_{AB}^{(0)}$ is fixed to be the unit sphere metric $\mathring q_{AB}$. We postpone the derivation of the general solution for any metric $g_{AB}^{(0)}$ to future endeavour.

We write $\xi^t_{(0)} \equiv f$ as in \eqref{dico diffeo}. The scalar field $\Phi$ and pseudo-scalar field $\Psi$ define the Helmholtz decomposition $\xi^A_{(0)} =\mathring q^{AB} \partial_B \Phi (t,x^C) + \mathring \epsilon^{AB}\partial_B \Psi(t,x^C)$. With our boundary condition $\sqrt{|g^{(0)}|} =  \sqrt{\mathring{q}}/ \ell$, one can show that 
\begin{equation}
\partial_t^2 f - \frac{\eta}{2\ell^2}D_A D^A f=0 \label{eq:WaveF}
\end{equation}
after taking a second derivative with respect to $t$ of the constraint equation for $f$. The evolution equation for $\xi^A_{(0)}$ is vectorial on the sphere, so we can solve for its curl and divergence separately. This yields respectively
\begin{align}
& D_A D^A (\partial_t \Psi) = \varepsilon_{AB} D^A W^B, \label{eq:WavePsi} \\
& D_A D^A \left( \partial_t \Phi -\frac{\eta}{\ell^2} f\right) = -D_A W^A, \label{eq:WavePhi}
\end{align}
where $W^A \equiv \partial_t g^{AB}_{(0)} \partial_B \Phi$ (at this stage, we have not yet imposed $g_{AB}^{(0)} = \mathring{q}_{AB}$). It is worth noticing that the wave equation \eqref{eq:WaveF} does not fully determine $f$, since this is a second order equation with respect to  $t$, while the original constraint on $f$ is a first order equation. The information we have lost by applying the second derivative is encoded in the remaining condition
\begin{equation}
\partial_t f = \frac{1}{2}D_A D^A \Phi
\label{eq:SupplCstf}
\end{equation}
that we need to take into account. We need to solve  \eqref{eq:WaveF}--\eqref{eq:SupplCstf}.

The equations for $\partial_t f$, $\partial_t\Phi$ and $\partial_t\Psi$ determine the fields $f$, $\Phi$ and $\Psi$ up to three arbitrary function of the angles $(\hat f,\hat \Phi,\hat \Psi)$. Let us show that these functions  are constrained. We obtain $D_A D^A \hat f = 0$ directly from \eqref{eq:WaveF} and $D_A D^A \hat \Phi = 0$ from \eqref{eq:SupplCstf}. Recalling that only constants are harmonic functions on compact manifolds, we can set $\hat f = f_0$ ($f_0\in\mathbb R$) and $\hat \Phi = 0$, since a constant value of $\Phi$ does not appear in the vector $\xi^A_{(0)}$. Hence we see that the solution for $\partial_t f$ and $\partial_t \Phi$ fully determines $f$ (up to a residual constant) and $\Phi$. However, we also observe that nothing constrains $\hat \Psi$, showing that it will remain an arbitrary function on the angles in $\Psi$. Finally the wave equation involving $f$ gives rise to two arbitrary functions of the angles as integration constants, which will be brought to $\Phi$ thanks to \eqref{eq:WavePhi}. Hence the number of arbitrary functions of $x^A$ is shown to be three. 

We now set $g^{(0)}_{AB} \equiv \mathring q_{AB}$ and $W^A$ vanishes identically. The solution of \eqref{eq:WavePsi} is $\partial_t\Psi = c$ for some real constant $c$. This constant can be removed as an ambiguity in defining the Helmholtz fields, since it is responsible for a linear term $ct$ in $\Psi$ which will never contribute to the actual vector $\xi^A_{(0)}$. The solution for $\Psi$ is thus simply $\Psi=\Psi(x^A)$. We can directly solve \eqref{eq:WaveF} by using Fourier transform methods. We obtain
\begin{equation}
f(t,x^A) = \int_0^\infty \text d\omega \, \left[ f_E (x^A) \cos \left(\frac{\omega \,t}{\ell}\right) + f_O(x^A) \sin \left(\frac{\omega \,t}{\ell}\right)\right]
\end{equation}
if $\Lambda <0$ and 
\begin{equation}
f(t,x^A) = \int_0^\infty \text d\omega \, \left[ f_E (x^A) \cosh \left(\frac{\omega \,t}{\ell}\right) + f_O(x^A) \sinh \left(\frac{\omega \,t}{\ell}\right)\right]
\end{equation}
if $\Lambda >0$. In both cases, the Fourier coefficients are constrained as follows:
\begin{equation}
\mathring D_A \mathring D^A f_{E} = -2\omega^2 f_{E},\quad \mathring D_A \mathring D^A f_{O} = -2\omega^2 f_{O}.
\end{equation}
These equations select discrete values for $\omega$ which satisfy $\omega_l^2 = \frac{1}{2} l(l+1)$ with $l\in\mathbb N$, and the solution for $f$ is given in terms of the real spherical harmonics $Y_{lm}(x^A)$ ($m\in\mathbb N$, $|m|\leq l$) as
\begin{equation}
f(t,x^A) = \left\lbrace
\begin{split}
\, & \sum_{l,m} \left[ a_{lm} \cos \left(\frac{\omega_l \,t}{\ell}\right) + b_{lm} \sin \left(\frac{\omega_l \,t}{\ell}\right)\right] Y_{lm} (x^A) \text{ if } \Lambda < 0,\\
\, & \sum_{l,m} \left[ a_{lm} \cosh \left(\frac{\omega_l \,t}{\ell}\right) + b_{lm} \sinh \left(\frac{\omega_l \,t}{\ell}\right)\right] Y_{lm} (x^A) \text{ if } \Lambda > 0,
\end{split}
\right. \label{eq:GeneralSolutionf}
\end{equation}
where $\lbrace a_{lm} \rbrace$ and $\lbrace b_{lm} \rbrace$ are two sets of real constants. From \eqref{eq:WavePhi}, we can deduce that
\begin{equation}
\partial_t\Phi - \frac{\eta}{\ell^2}f = 0 \label{eq:PartialPhi}
\end{equation}
up to some real constant that can again be removed as an ambiguity in defining $\Phi$. Taking one more derivative with respect to $t$ and recalling that \eqref{eq:SupplCstf} holds, we obtain that $\Phi$ satisfies the same wave equation as $f$,
\begin{equation}
\partial_t^2 \Phi - \frac{\eta}{2\ell^2}\mathring D_A \mathring D^A \Phi=0. \label{eq:BoxPhi}
\end{equation}
The general solution has already been derived (see \eqref{eq:GeneralSolutionf}), and reads as \eqref{eq:PhiLBMS4}, involving two new sets of real constants $\lbrace A_{lm}\rbrace$ and $\lbrace B_{lm}\rbrace$. The $1/\ell$ factor in front of the $B$'s is for now purely conventional, but will ensure later that the coefficients $A_{lm}$ and $B_{lm}$ will not depend on $\ell$ in the flat limit process $\ell\to \infty$. These new sets of constants are not independent of $\lbrace a_{lm}\rbrace$ and $\lbrace b_{lm}\rbrace$, since we must require that the remaining constraints \eqref{eq:SupplCstf} and \eqref{eq:PartialPhi} hold. The first one has been implemented in the derivation of the wave equation \eqref{eq:BoxPhi}, so we just have to impose the second one. This yields
\begin{equation}
a_{lm} = \eta\,B_{lm}\,\omega_l, \quad b_{lm} = -\ell \,A_{lm}\,\omega_l.
\end{equation}
Hence, the $f$ gauge parameter reads as \eqref{eq:fLBMS4}.

The exact isometries of global (A)dS$_4$ are recovered if we restrict ourselves to the lowest modes $l=0$ and $l=1$. Indeed, if one requires further that $\delta_{\xi} g_{AB}^{(0)} = 0$, it comes
\begin{equation}
\mathring D_A \xi^{(0)}_B+\mathring D_B \xi^{(0)}_A-\mathring D_C \xi_{(0)}^C \mathring q_{AB}=0 \Leftrightarrow \left\lbrace 
\begin{split}
(\mathring D_A \mathring D_B \Phi)^{TF} &= 0, \\
\varepsilon_{C(A}\mathring D_{B)}\partial^C \Psi &= 0.
\end{split}
\right.
\end{equation}
In stereographic coordinates $(z,\bar z)$, if we introduce the auxiliary fields $\phi \equiv (1+z\bar z)\Phi $ and $\psi \equiv (1+z\bar z)\Psi $, these equations become simply $\partial_z^2 \phi = 0 = \partial_{\bar z}^2 \phi$ and $\partial_z^2 \psi = 0 = \partial_{\bar z}^2 \psi$. Hence $\phi$ and $\psi$ are at most linear in $z$ and $\bar z$, the only non-linear piece that can appear being the squared modulus $z\bar z$. In conclusion, the solution to the conformal Killing equation developed for the Helmholtz fields $\Phi$ and $\Psi$ only involve the lowest (real) spherical harmonics with $l=0,1$.

\section{Useful relations for computations in Bondi gauge}
\label{Useful relations}

In this Appendix, we give some useful relations  in Bondi gauge that are widely used in Section \ref{Flat limit of}. The metric on the celestial sphere is written as $q_{AB}$ and we always impose that $\delta\sqrt{q}=0$. From $\delta q^{AB} = -q^{AC}q^{BD} \delta q_{CD}$, we have for any symmetric tensor $T_{AB}$
\begin{equation}
T_{AB} \delta q^{AB} = - T^{AB}\delta q_{AB}. \label{prop1}
\end{equation} For any $C_{AB}$ symmetric traceless tensor (\textit{i.e.} $C_{AB}=C_{(AB)}$, $q^{AB}C_{AB}=0$), it follows that 
\begin{equation}
q_{AB} \delta C^{AB} = C_{AB} \delta q^{AB}.
\end{equation} Using explicitly $\delta\sqrt{q}=0$ one finds
\begin{align}
\delta (C_{AB} C^{AB}) &= 2 C_{AB}\delta C^{AB}, \\
\delta (C_{AB} C^{AB}) &= q_{AB} \delta (C^{AC} C_C^B ), \\
M_{AB} \delta C^{AB} &= M^{AB} \delta C_{AB}, \label{prop5}
\end{align}
where $M_{AB}$ is also an arbitrary symmetric traceless tensor. Considering two variations $\delta,\delta'$ one can prove the following identity, again with $\delta\sqrt{q}=0$ :
\begin{equation}
T_{AB} \delta\delta' q^{AB} = - T^{AB} \delta \delta' q_{AB} - T \delta q_{AB} \delta' q^{AB}, 
\end{equation} where $T = q^{AB} T_{AB}$.
It follows from (\ref{prop1}) that
\begin{equation}
\delta T_{AB} \delta' q^{AB} = - \delta T^{AB} \delta' q_{AB} + T \delta q_{AB} \delta' q^{AB}. \label{prop9}
\end{equation}
For a traceless tensor such as $M_{AB}$, we have
\begin{equation}
\delta M_{AB} \delta' q^{AB} = - \delta M^{AB} \delta' q_{AB}.
\end{equation}
For the metric $q_{AB}$ itself, $T=2$ and
\begin{equation}
\delta q_{AB} \delta' q^{AB} = \delta q^{AB} \delta' q_{AB}.
\end{equation}
Finally from (\ref{prop5}) we get
\begin{equation}
\delta M_{AB} \delta' C^{AB} = \delta M^{AB}\delta' C_{AB}. \label{eq:DeltaSymCC}
\end{equation}

\section{Flat limit of solution space and symmetries}
\label{Flat limit of solution space and symmetries}

In this Appendix, we recall some key results of \cite{Compere:2019bua} regarding the flat limit process in Bondi gauge (see also \cite{Ruzziconi:2019pzd} for a review).

\subsection{Flat limit of the solution space}
\label{app Flat limit of the solution space}

 Consider Bondi gauge \eqref{Bondi metric} with the boundary conditions \eqref{eq:gABFallOff} and \eqref{LBMS COND BONDI}. The solution space for the Einstein equations with non-vanishing cosmological constant ($\Lambda \neq 0$) is decribed firstly by \eqref{solution space fall offs}, with the evolution constraints with respect to the $u$ coordinate \eqref{eq:EvolutionEquations}. Secondly, it is described  in \eqref{eq:gABFallOff}, taking the constraints \eqref{conseq det} ,\eqref{equation no log} and \eqref{eq:CAB} into account (in particular, as discussed in \cite{Compere:2019bua}, all the terms of $\mathcal{O}(r^{-2})$ in \eqref{eq:gABFallOff} are completely determined in terms of $q_{AB}$ and $\mathcal{E}_{AB}$). In summary, the data parametrizing this solution space is given by
\begin{equation}
\{ q_{AB}, \mathcal{E}_{AB}, M, N_A \}_{\Lambda \neq 0} , 
\label{sol ads}
\end{equation} where $M$ and $N_A$ satisfy the equations given in \eqref{eq:EvolutionEquations}. In other words, the characteristic initial value problem is completely specified by giving $q_{AB}(u,x^C$), $\mathcal{E}_{AB}(u,x^C)$, $M(u_0,x^C)$, $N_A(u_0, x^C)$, where $u_0$ is an initial value of $u$. 

Now, if we consider the flat limit of this solution space following the prescription recalled in Section \ref{secflat}, we get exactly the analytic part of the solution space in asymptotically flat spacetime considered in \cite{Compere:2018ylh} (the latter was initially obtained by solving the Einstein equations in Bondi gauge with $\Lambda = 0$ and assuming the boundary conditions \eqref{eq:gABFallOff}-\eqref{LBMS COND BONDI} and analyticity in the powers of $r$ expansions). Indeed, the left-hand side of equation \eqref{eq:CAB} goes to zero in the limit $\Lambda \to 0$, which means that the asymptotic shear $C_{AB}$ becomes unconstrained and $q_{AB}$ gets an evolution equation with respect to the $u$ coordinate
\begin{equation}
\partial_u q_{AB} = 0.
\end{equation} Similarly, equation \eqref{equation no log} is trivially satisfied in the limit and does not impose any constraint on $\mathcal{D}_{AB}$. However, as in \cite{Compere:2018ylh}, we will assume $\mathcal{D}_{AB} = 0$ for analyticity requirements. Furthermore, as discussed in \cite{Compere:2019bua}, the same phenomenon as in \eqref{eq:CAB} happens for the traceless parts of the subleading terms $\mathcal{O}(r^{-1})$ in the expansion \eqref{eq:gABFallOff}, namely they become free data of the solution space, but with fixed evolution with respect to $u$. Now, taking $\Lambda \to 0$ on the expansions \eqref{solution space fall offs} leads trivially to 
\begin{equation}
\begin{split}
\beta &= \frac{1}{r^2} \left[- \frac{1}{32} C^{AB} C_{AB}   \right] + \mathcal{O}(r^{-4}), \\
U^A &= \frac{1}{r^2} \left[ - \frac{1}{2} D_B C^{AB} \right] + \frac{1}{r^3} \left[ -\frac{2}{3} N^A + \frac{1}{3} C^{AB} D^C C_{BC} \right] + \mathcal{O}(r^{-4}), \\
\frac{V}{r} &= - \frac{R[q]}{2} - \frac{2M}{r} + \mathcal{O}(r^{-2}),
\end{split}
\end{equation} where $M$ and $N_A$ satisfy evolution equations with respect to $u$ that are obtained by taking the flat limit of \eqref{eq:EvolutionEquations} following carefully the recipe given in Section \ref{secflat}. This yields
\begin{equation}
\begin{split}
\partial_u   N_A &- \partial_A  M - \frac{1}{4} C_{AB} \partial^B R[q] - \frac{1}{16} \partial_A (N_{BC}C^{BC})  \\
& +\frac{1}{4} N_{BC} D_A C^{BC} + \frac{1}{4} D_B (C^{BC} N_{AC} - N^{BC} C_{AC})  \\
&+\frac{1}{4} D_B (D^B D^C C_{AC} - D_A D_C C^{BC}) = 0,
\end{split}
\end{equation} and 
\begin{equation}
\begin{split}
\partial_u  M +\frac{1}{8} N_{AB} N^{AB} - \frac{1}{8} D_A D^A R[q] 
- \frac{1}{4} D_A D_B N^{AB} = 0. 
\end{split}\label{duM}
\end{equation} In summary, the solution space obtained in the flat limit is the one of \cite{Compere:2018ylh} and is parametrized by the following data\footnote{The ``$\ldots$'' in \eqref{flat solution space} denotes an infinite tower of symmetric traceless two-dimensional tensors coming from the $\mathcal{O}(r^{-2})$ of the expansion \eqref{eq:gABFallOff} and that satisfy evolution equations with respect to the $u$ coordinate.} 
\begin{equation}
\{ q_{AB}, C_{AB}, M, N_A, \mathcal{E}_{AB}, \ldots \}_{\Lambda = 0} ,
\label{flat solution space}
\end{equation} where $q_{AB}$, $M$, $N_A$, $\mathcal{E}_{AB}$, $\ldots$ satisfy evolution equations with respect to $u$.  In other words, the characteristic initial value problem is completely specified by giving $C_{AB}(u,x^C)$, $q_{AB}(u_0,x^C)$, $M(u_0,x^C)$, $N_A(u_0, x^C)$, $\mathcal{E}_{AB}(u_0,x^C)$, $\ldots$ where $u_0$ is an initial value of $u$.

Notice that in all this flat limit process, we assumed that the functions \eqref{sol ads} parametrizing the solution space for $\Lambda \neq 0$ do no depend on $\Lambda$ (see \textit{e.g.} \cite{Campoleoni:2018ltl} for an example in three-dimensional gravity where this condition is relaxed). 

\subsection{Flat limit of the symmetries}
\label{flatlimits}

Let us now start from the $\Lambda$-BMS$_4$ asymptotic Killing vectors, which are given by the residual gauge diffeomorphisms \eqref{eq:xir} where the the parameters $f$, $Y^A$ and $\omega$ satisfy the constraints equations \eqref{constraints residuals}. Using the adjusted Lie bracket \eqref{eq:ModLieBracket}, they satisfy the commutation relations 
\begin{equation}
[\xi(f_1 , Y^A_1), \xi(f_2 , Y^A_2)]_\star = \xi(\hat{f} , \hat{Y}^A) ,
\label{struc const 2}
\end{equation} where 
\begin{equation}
\begin{split}
\hat{f} &= Y^A_1 \partial_A f_2 + \frac{1}{2} f_1  D_A Y^A_2 - \delta_{\xi(f_1 , Y^A_1)}f_2 - (1 \leftrightarrow 2 ), \\
\hat{Y}^A &= Y_1^B \partial_B Y^A_2 - \frac{\Lambda}{3} f_1 q^{AB} \partial_B f_2 - \delta_{\xi(f_1 , Y^A_1)} Y^A_2 - (1 \leftrightarrow 2 ). \label{structure constant 2}
\end{split}
\end{equation} which is a simple translation of \eqref{struc const} and \eqref{structure constant} using \eqref{dico diffeo}. These can also be derived direclty in Bondi gauge following the procedure described in \cite{Barnich:2010eb}  (see \cite{Ruzziconi:2019pzd}). 

Now, taking the flat limit $\Lambda \to 0$, the form of the asymptotic Killing vectors \eqref{eq:xir} is the same, except that they are evaluated on the flat solution space parametrized by \eqref{flat solution space} instead of the one parametrized by \eqref{sol ads}, while the parameters $f$, $Y^A$ and $\omega$ satisfy from now on
\begin{equation}
\partial_u f = \frac{1}{2} D_A Y^A , \qquad \partial_u Y^A = 0, \qquad \omega = 0.
\label{eq param bond}  
\end{equation} Notice that these equations do not involve the solution space. Therefore, the parameters are field-independent in the flat limit ($ \delta_{\xi(f_1 , Y^A_1)}f_2 = 0 = \delta_{\xi(f_1 , Y^A_1)} Y^A_2$). The equations \eqref{eq param bond} can be readily solved as
\begin{equation}
f = T(x^A) + \frac{1}{2} u D_A Y^A, \qquad Y^A = Y^A(x^B), 
\label{solut fY}
\end{equation} where $T$ and $Y^A$ are the supertranslation and superrotation generators, respectively. The flat limit of the commutation relations \eqref{struc const 2} and \eqref{structure constant 2} is straightforward and yields 
\begin{equation}
[\xi(f_1 , Y^A_1), \xi(f_2 , Y^A_2)]_\star = \xi(\hat{f} , \hat{Y}^A) ,
\label{struc const 3}
\end{equation} where 
\begin{equation}
\begin{split}
\hat{f} &= Y^A_1 \partial_A f_2 + \frac{1}{2} f_1  D_A Y^A_2 - (1 \leftrightarrow 2 ), \\
\hat{Y}^A &= Y_1^B \partial_B Y^A_2 - (1 \leftrightarrow 2 ). \label{structure constant 3}
\end{split}
\end{equation}  Using \eqref{solut fY}, the commutation relations \eqref{struc const 3} and \eqref{structure constant 3} can be rewritten as
\begin{equation}
[\xi(T_1 , Y^A_1), \xi(T_2 , Y^A_2)]_\star = \xi(\hat{T} , \hat{Y}^A) ,
\label{struc const 4}
\end{equation} where 
\begin{equation}
\begin{split}
\hat{T} &= Y^A_1 \partial_A T_2 + \frac{1}{2} T_1  D_A Y^A_2 - (1 \leftrightarrow 2 ), \\
\hat{Y}^A &= Y_1^B \partial_B Y^A_2 - (1 \leftrightarrow 2 ). \label{structure constant 4}
\end{split}
\end{equation} These are precisely the commutation relations of the (generalized) BMS$_4$ asymptotic symmetry algebra given by the semi-direct sum Diff($S^2$)$\loplus \mathcal{S}$, where Diff($S^2$) are the superrotations parametrized by $Y^A$ and $\mathcal{S}$ are the supertranslations parametrized by $T$ \cite{Campiglia:2014yka , Campiglia:2015yka, Compere:2018ylh , Flanagan:2019vbl , Campiglia:2020qvc}.

Finally, let us show that the flat limit obtained by taking directly $\ell\to\infty$ in \eqref{eq:PhiLBMS4} and \eqref{eq:fLBMS4} reproduces the expression of the (generalized) BMS$_4$ generators \eqref{solut fY}. Since the flat limit of the phase space is well-defined within the Bondi gauge, we now consider $f$, $\Phi$ and $\Psi$ defined in the Appendix \ref{app:LBMS4gen} as functions of $(u,x^A)$, where $u$ is the Bondi retarded time and $x^A$ remain the angular coordinates. We see that the constraint equations \eqref{eq:LBMS4constraints} that we solved in this Appendix are unchanged by the diffeomorphism between Starobinsky/Fefferman-Graham and Bondi coordinates described in Appendix B of \cite{Compere:2019bua}, up to the natural replacement of $t$ by $u$ (see equations \eqref{constraints residuals}). We also recall that the residual gauge parameters are related as \eqref{dico diffeo}. The flat limit of \eqref{eq:PhiLBMS4} gives
\begin{equation}
\Phi(u,x^A) = \sum_{l,m} A_{lm} Y_{lm}(x^A) = \Phi(x^A)
\end{equation}
as expected. The same limit of \eqref{eq:fLBMS4} leads to
\begin{equation}
\begin{split}
f(u,x^A) &= \sum_{l,m} \left[ \eta B_{lm} - \omega_l \, u\, A_{lm} \right] \omega_l \, Y_{lm}(x^A) \\
&= \sum_{l,m} \left[ \tilde B_{lm} \, Y_{lm}(x^A) - \omega_l^2 \, u\, A_{lm} \, Y_{lm}(x^A) \right], \, \left(\tilde B_{lm} \equiv \eta \, \omega_l B_{lm}\right) \\
&= \sum_{l,m} \tilde B_{lm} \, Y_{lm}(x^A) + \frac{u}{2}\sum_{l,m} A_{lm} \mathring D_B \mathring D^B Y_{lm}(x^A) \\
&= T(x^A) + \frac{u}{2}\mathring D^B \mathring D_B \Phi(x^A) \\
&= T(x^A) + \frac{u}{2} \mathring D_B Y^B(x^A),
\end{split}
\end{equation}
where $T(x^A)$ is an arbitrary scalar field on the celestial sphere. Hence we recovered \eqref{solut fY}.

\providecommand{\href}[2]{#2}\begingroup\raggedright\endgroup

\end{document}